\documentclass{emulateapj}

\newcommand{\be}{\begin{equation}}
\newcommand{\ee}{\end{equation}}
\newcommand{\beqn}{\begin{eqnarray}}
\newcommand{\eeqn}{\end{eqnarray}}
\newcommand{\bld}[1]{\mbox{\boldmath$#1$\unboldmath}}
\newcommand{\mycite}[1]{\citeauthor{#1}\ \citeyear{#1}}

\begin{document}

\title{Formation, Survival, and Destruction of Vortices in Accretion Disks}

\author{Yoram Lithwick\altaffilmark{1}}
\altaffiltext{1}{CITA. Toronto, Ontario, Canada; yoram@cita.utoronto.ca}

\begin{abstract}

Two dimensional hydrodynamical disks
 are nonlinearly unstable to the formation of vortices.  
 Once formed, these vortices essentially survive forever.  
 What happens in three dimensions?
 We show with incompressible shearing box simulations that  in 3D a vortex in
 a short box forms and survives just as in 2D.  But a vortex
 in a tall box is unstable and is destroyed. 
In our simulation, the unstable vortex decays into a transient turbulent-like state that 
transports angular momentum outward at a nearly constant rate
for hundreds of orbital times.
The 3D  instability that destroys vortices is a generalization of  the 2D instability
that forms them.
We derive  the conditions for these nonlinear instabilities to act
by calculating the coupling between linear modes, and
thereby derive the criterion
for a vortex to survive in 3D as it does in 2D: {\it the azimuthal extent of the
vortex must be larger than the scale height  of the accretion disk}. 
 When this criterion is violated, the
vortex is unstable and decays.
Because vortices are longer in azimuthal than
in radial extent by a factor that is inversely proportional to their
excess vorticity, a vortex with given radial extent will only survive in a 3D disk
if it is sufficiently weak.  This counterintuitive result explains why previous
3D simulations always yielded decaying vortices: their 
vortices were too strong.
 Weak vortices behave two-dimensionally even if their width is much less than their height
 because they are stabilized by rotation, and behave as Taylor-Proudman columns.
We conclude that in protoplanetary disks  weak vortices can trap dust and serve
as the nurseries of planet formation.  Decaying strong vortices might be responsible
for the outwards transport of angular momentum that is required to make accretion
disks accrete.

 \end{abstract}

\section{Introduction}

Matter accretes onto a wide variety of objects, such as young stars, black holes, and white
dwarfs, through accretion disks.  In highly ionized disks  magnetic fields are important, 
and they trigger turbulence via 
the magnetorotational instability \citep{BH98}.  However, many disks,
 such as those around
young stars or dwarf novae, are nearly neutral  \citep[e.g.,][]{SMUN00,GM98}. 
In these disks, 
the fluid motions
are well described by hydrodynamics. 

Numerical simulations of hydrodynamical 
disks in two-dimensions---in the plane of the disk---often
produce long-lived vortices \citep{GL99,UR04,JG05}.
If vortices really exist in accretion disks, 
they can have important consequences.  First and foremost,
they might generate turbulence.
Since turbulence naturally transports angular momentum outwards\footnote{
Energy conservation implies that turbulence transports angular momentum outwards;
see \S \ref{sec:pseudo}.  
 Nonetheless, if an external energy source  (e.g., the radiative
energy from the central star) drives the turbulence, then angular momentum could in principle be transported 
inwards.
}, 
as is required for mass to fall inwards, it might be vortices 
that cause accretion disks to accrete.
Second, in disks around young stars, long-lived vortices can trap solid particles
and initiate the formation of planets \citep{BS95}.

Why do vortices naturally form in 2D simulations?
Hydrodynamical disks are stable to linear perturbations.  
However, they are nonlinearly unstable, despite some claims to the 
contrary in the
astrophysical literature.
In two dimensions, the incompressible hydrodynamical
equations of a disk are equivalent to those of a non-rotating linear
shear flow \citep[e.g.,][hereafter L07]{L07}.
And it has long been known that such flows are nonlinearly unstable 
(\mycite{Gill65}; \mycite{LK88}; L07).  This nonlinear instability is just 
a special case of the Kelvin-Helmholtz instability. 
Consider a linear shear flow extending throughout the $x$-$y$ plane
with velocity profile $\bld{v}=-qx\bld{\hat{y}}$, where $q>0$ 
is the constant shear rate, so that $-q$ is the flow's vorticity.
(In the equivalent accretion disk, the local angular speed is 
$\Omega=2q/3$.)  This shear flow is linearly stable to 
infinitesimal perturbations.  But if the shear profile is altered
by a small amount, the alteration can itself be unstable
to infinitesimal perturbations.  To be specific, let the alteration
be confined
within a band of width $\Delta x$, and let it 
have vorticity $\omega=\omega(x)$ (with $|\omega|\lesssim q$), so that it induces a velocity field
in excess of the linear shear
with components $u_y\sim \omega\Delta x$ and $u_x=0$.  
Then this band is unstable to infinitesimal nonaxisymmetric
(i.e. non-stream-aligned) perturbations provided roughly that
\be
\left|k_y
\right|
\lesssim
{1\over q}{|\omega|\over \Delta x}
\ \ \Rightarrow 
{\rm 2D\ instability}
\label{eq:2dinst}
\ee
where $k_y$ is the wavenumber of the nonaxisymmetric perturbation.\footnote{
More precisely, the necessary and sufficient condition for instability
in the limit $|\omega|\ll q$ is that $|k_y|<{1\over 2 q}\int_{-\infty}^\infty {d\omega/dx\over x-x_0}dx$, where
$x_0$ is any value of $x$ at which $d\omega/dx=0$ \citep[][L07]{Gill65,LK88}.
For arbitrarily large  $\omega$, 
Rayleigh's inflection point theorem and Fj\o rtoft's theorem give necessary (though
insufficient) criteria for instability \citep{DR04}.  The former states that for 
instability, it is required that $d\omega/dx=0$ somewhere in the flow, i.e. that the velocity
field must have an inflection point.  \cite{Love99} generalize Rayleigh's inflection
point theorem to compressible and nonhomentropic disks.
\label{foot:inst}
}
For any value of $|\omega|$ and $\Delta x$, the band is always unstable 
to perturbations with long enough wavelength.
Remarkably, instability even occurs when $|\omega|$ is infinitesimal.
Hence we may regard this  as a true nonlinear instability.  
\cite{BH06} assert that detailed numerical simulations have not shown evidence for nonlinear
instability. The reason many simulations fail to see it
is that their boxes are not long enough in the $y$-direction to encompass a small
enough non-zero $|k_y|$.

In two dimensions, the outcome of this instability is a long-lived vortex (e.g., L07).
A vortex that has been studied in detail is the Moore-Saffman vortex, 
which is a localized patch of spatially
constant vorticity  superimposed on a linear shear flow  \citep{Saffman95}.
When $|\omega|\lesssim q$, 
where $\omega$ here refers to the spatially constant
excess vorticity within the patch, and when the vorticity within
the patch ($\omega-q$) is stronger than that of the background shear,
then the patch forms a stable vortex that is elongated in $y$ relative
to $x$ by the factor 
\be
{\Delta y\over \Delta x} \sim {q\over |\omega|} \ . \label{eq:ms}
\ee
This relation applies not only to Moore-Saffman vortices,
but also to vortices whose $\omega$ is not spatially constant.
It may be understood as follows.
A patch with characteristic excess vorticity $\sim\omega$ and with $\Delta y\gg \Delta x$
induces a velocity field in the $x$-direction with amplitude $u_x\sim|\omega|\Delta x$, 
independent of the value of $\Delta y$ (e.g., \S 6 in L07). As long as $|\omega|\lesssim q$,
the $y$-velocity within the vortex is predominantly due to the background shear, 
and is $\sim q\Delta x$. Therefore the time to cross the width of the vortex is
$t_x\sim \Delta x/u_x \sim 1/|\omega|$,
and the time to cross its length is 
$t_y\sim \Delta y/(q\Delta x)$.  Since these times must be comparable in a vortex,
equation (\ref{eq:ms}) follows.
Equation (\ref{eq:ms}) is very similar to equation (\ref{eq:2dinst}). 
The 2D instability
naturally forms into a 2D vortex.  
Futhermore, the exponential growth rate of the instability is $\sim |\omega|$, 
which is comparable to the rate at which fluid circulates around the vortex.

More generally,
an arbitrary axisymmetric profile of $\omega(x)$ tends to evolve into a 
distinctive
banded structure.  Roughly speaking, bands where $\omega<0$ contain vortices, and these
are interspersed 
with bands where $\omega>0$, which  contain no vortices.   (Recall that we take
the background vorticity to be negative; otherwise, the converse would hold.)
The reason for this  is that only regions that have $\omega<0$ can be unstable,
 as may be inferred either from the integral
criterion for instability given in footnote \ref{foot:inst}, or from Fj\o rtoft's theorem.
For more detail on vortex dynamics in shear flows, see
the review by \cite{Marcus93}.

What happens in three dimensions? To date, numerical simulations of 
vortices in 3D disks have been reported in two papers.
\cite{BM05b} 
initialized their simulation with a Moore-Saffman vortex, and
solved the anelastic equations in a stratified disk.
They found that this vortex decayed.  As it decayed,
new vortices were formed
in the disk's atmosphere, two scale heights above the midplane. The new vortices
 survived for the duration of the simulation.
\cite{SSG06} performed both 2D and 3D simulations of the compressible 
hydrodynamical equations in an unstratified disk, initialized with 
large random fluctuations.  They found that whereas the 2D simulations
produced long-lived vortices, 
in three dimensions  vortices rapidly decayed.

 Intuitively, it seems
clear that a vortex in a very thin disk will behave as it does in  2D.
And from the  3D simulations described above it may be inferred that placing this vortex in a
very thick disk will induce its decay. 
Our main goal in this paper is to understand  these two behaviors, and the transition
between them.
A crude explanation of our final result is that vortices decay when the 2D
vortex motion
couples resonantly to 3D modes, i.e., to modes that have vertical wavenumber
$k_z\ne 0$.  As described
above, a vortex with excess vorticity $|\omega|$ has circulation frequency $\sim|\omega|$, 
and $k_y/k_x\sim |\omega|/q$, where $k_x$ and $k_y$ are its ``typical'' wavenumbers.
Furthermore, it is well-known that the frequency of axisymmetric ($k_y=0$) inertial
waves is $\Omega k_z/\sqrt{k_x^2+k_z^2}$  (see eq.  [\ref{eq:axi3d}]).
Equating the two frequencies, and taking the $k_x$ of the 3D mode to be comparable
to the $k_x$ of the vortex, as well as setting $q=3\Omega/2$ for a Keplerian disk, we find
\be
k_z\sim k_y \label{eq:res}
\ee
as the condition for resonance.
Therefore a vortex with length $\Delta y$  will survive in a box
with height $\Delta z\lesssim \Delta y$, because in such a box
all 3D modes have too high a frequency to couple with the vortex, i.e., all
 nonzero $k_z$
exceed the characteristic $k_y\sim 1/\Delta y$.
But when $\Delta z\gtrsim \Delta y$, there exist $k_z$ in the box that satisfy
the resonance condition (\ref{eq:res}), leading to the vortex's destruction.
This conclusion suggests that vortices live indefinitely in disks with scale height
less than their length ($h\lesssim \Delta y$)
because in such disks all 3D modes have too high a frequency for resonant coupling.
This conclusion is also consistent  with the simulations of
\cite{BM05b} and \cite{SSG06}. Both of these works
 initialized their simulations with strong excess vorticity
$|\omega|\sim q$, corresponding to nearly circular vortices.  
Both had vertical domains that were comparable to the vortices' width.
Therefore both saw that their vortices decayed.  Had they initialized their simulations
with smaller $|\omega|$, and increased the box length $L_y$ to encompass
the resulting elongated vortices, both would have found long-lived 3D vortices.
 \citeauthor{BM05b}'s discovery of long-lived vortices in the disk's
atmosphere is simple to understand because  the local scale height is reduced
in inverse proportion to the height above the midplane.   Therefore 
higher up in the atmosphere the dynamics becomes more two-dimensional, and
a given vortex is better able to survive the higher it is.\footnote{
However, \cite{BM05b} also include buoyancy forces in their simulations,
which we ignore here.  How buoyancy affects the stability of vortices is a topic
for future work.
}

\subsection{Organization of the Paper}

In  \S \ref{sec:eom} we introduce the equations of motion, and
in \S \ref{sec:pseudo} we present
 two pseudospectral simulations.
 One illustrates the
formation and survival of a vortex in a short box, and the other illustrates
the destruction of a vortex in a tall box. 

In \S\S \ref{sec:lin}-\ref{sec:nonlin} we develop a theory explaining
this behavior.
The reader who is satisfied by the qualitative description leading
to equation (\ref{eq:res}) may
skip those two sections.
The theory that we develop is indirectly related to the transient
amplification scenario for the generation of turbulence.
 Even though 
hydrodynamical disks are linearly stable, linear perturbations 
can be transiently amplified before they decay, often by a 
large factor.  It has been proposed that sufficiently amplified modes might
couple nonlinearly, leading to turbulence \citep[e.g.,][]{CZTL03,Yecko04,AMN05}.
However, to make this proposal more concrete, one must work out
how modes couple nonlinearly.  In  L07, we did that in two dimensions.  
We showed that the 2D nonlinear instability of equation (\ref{eq:2dinst}) is a consequence
of the coupling of an axisymmetric mode with a transiently amplified mode, 
which may be called a ``swinging mode'' because its phasefronts are swung around
by the background shear. 
In \S \ref{sec:nonlin} we show that the 3D instability responsible
for the destruction of vortices is a generalization of this 2D instability. 
It may be understood by examining the coupling of a 
3D swinging mode 
with an axisymmetric one.    3D modes become increasingly unstable as $|k_z|$
decreases, and in the limit that $k_z\rightarrow 0$, the 3D instability matches
smoothly onto the 2D one.
Thicker disks are more prone to 3D instability because they encompass
smaller $|k_z|$.

\section{Equations of Motion}
\label{sec:eom}

We solve the  ``shearing box'' equations,
which approximate the dynamics in an accretion disk on lengthscales
much smaller than the distance to the disk's center. 
We assume incompressibility, which is a good approximation
when relative motions are subsonic.
We also neglect vertical gravity, and hence stratification and buoyancy, which is 
an oversimplification. 
To fully understand vortices in astrophysical disks, one must consider the effects
of  vertical gravity as well as of shear and rotation. 
In this paper, we consider only two pieces of this puzzle---shear and
rotation.  Adding the third piece---vertical gravity---is a topic that 
we leave for future investigations.  See also the Conclusions for some 
speculations.

An unperturbed Keplerian disk has angular velocity profile $\Omega(r)\propto r^{-3/2}$.
In a reference
frame rotating at constant angular speed $\Omega_0\equiv \Omega(r_0)$, where $r_0$ is
a fiducial radius, 
the incompressible shearing box equations of motion  read
\begin{eqnarray}
{\partial_t {\bld v}}+{\bld{ v\cdot\nabla v}}&=&-
2\Omega_0{\bld{\hat z}}{\bld{\times v}}+2q\Omega_0x{\bld{ \hat{x}}} -{\bld\nabla} P/\rho\ \ ,
\label{eq:bigeom}
\\
\bld{\nabla\cdot v}&=&0  
\end{eqnarray}
adopting
 Cartesian coordinates $x,y,z$, which are related to the disk's cylindrical $r,\theta$ 
via $x\equiv r-r_0$ and $y\equiv r_0(\theta-\Omega_0t)$;
$\bld{\hat{x}}$ and $\bld{\hat{z}}$ are unit vectors, 
and
\be
q\equiv -{d \Omega\over d\ln r}\Big\vert_{r_0}={3\over 2}\Omega_0 \ ,
\ee
We retain $q$ and $\Omega_0$ as independent parameters because they 
parameterize different effects: shear and rotation, respectively. 
The first term on the right-hand side of equation (\ref{eq:bigeom}) is the Coriolis force, 
and the second is what remains after adding centrifugal and gravitational forces.
Decomposing  the velocity  into
\be
\bld{v}=-qx\bld{\hat{y}}+\bld{u} \ , \label{eq:dec}
\ee
where the first term is the shear flow of the unperturbed disk,  yields
\begin{eqnarray}
\left( {\partial_t}-qx{\partial_y}\right) {\bld u}
+{\bld{u\cdot\nabla u}}
&=&
 q u_x\bld{\hat{y}}
 -2\Omega {\bld{\hat z}}{\bld{\times u}}
 -\bld{\nabla} P/\rho \label{eq:eom2}
\\
{\bld{\nabla\cdot u}} &=&
0   \  ,
\label{eq:eom1}
\end{eqnarray}
dropping the subscript from $\Omega_0$, as we shall do 
in the remainder of this paper.  An unperturbed disk has 
$\bld{u}=0$.

In addition to the above ``velocity-pressure'' formulation, an
alternative ``velocity-vorticity'' formulation will prove convenient.  It
is given by 
the curl of equation (\ref{eq:eom2}), 
\be
(\partial_t-qx\partial_y)\bld{\omega}=-q\bld{\hat{y}}\omega_x+(2\Omega-q)\partial_z\bld{u}+
\bld{\nabla\times}(\bld{u\times \omega})
\label{eq:eomo}
\ee
where
\be
\bld{\omega}\equiv \bld{\nabla\times u}
\label{eq:omdef}
\ee
is the vorticity of $\bld{u}$.
 Equation (\ref{eq:eomo}),  together with the inverse of equation (\ref{eq:omdef})
\be
\bld{u}=-\nabla^{-2}\bld{\nabla\times\omega} \label{eq:ominv} \ ,
\ee
form a complete set. 

Equation (\ref{eq:eomo}) implies that the total vorticity field
is frozen into the fluid, because it is equivalent to 
\be
\label{eq:omtot}
\partial_t\bld{\omega}_{\rm tot}=\bld{\nabla\times}(\bld{v\times\omega}_{\rm tot}) \ ,
\ee
where
\be
{\bld \omega}_{\rm tot}\equiv (2\Omega-q)\bld{\hat{z}}+\bld{\omega} \ 
\ee
is the total vorticity; note that
 $-q\bld{\hat{z}}$ is the vorticity of the unperturbed shear flow in the rotating frame, 
and hence $(2\Omega-q)\bld{\hat{z}}$ is the unperturbed vorticity in the non-rotating frame.
The vorticity-velocity picture is similar to MHD, where it is the
magnetic field that is frozen-in because it satisfies equation (\ref{eq:omtot})
in place of $\bld{\omega}_{\rm tot}$.  However, in MHD the velocity field has its own dynamical
equation, whereas in incompressible hydrodynamics it is determined directly from the vorticity field
via equation (\ref{eq:ominv}).

\begin{figure*}
  \centering
  \includegraphics[width=.32\textwidth]{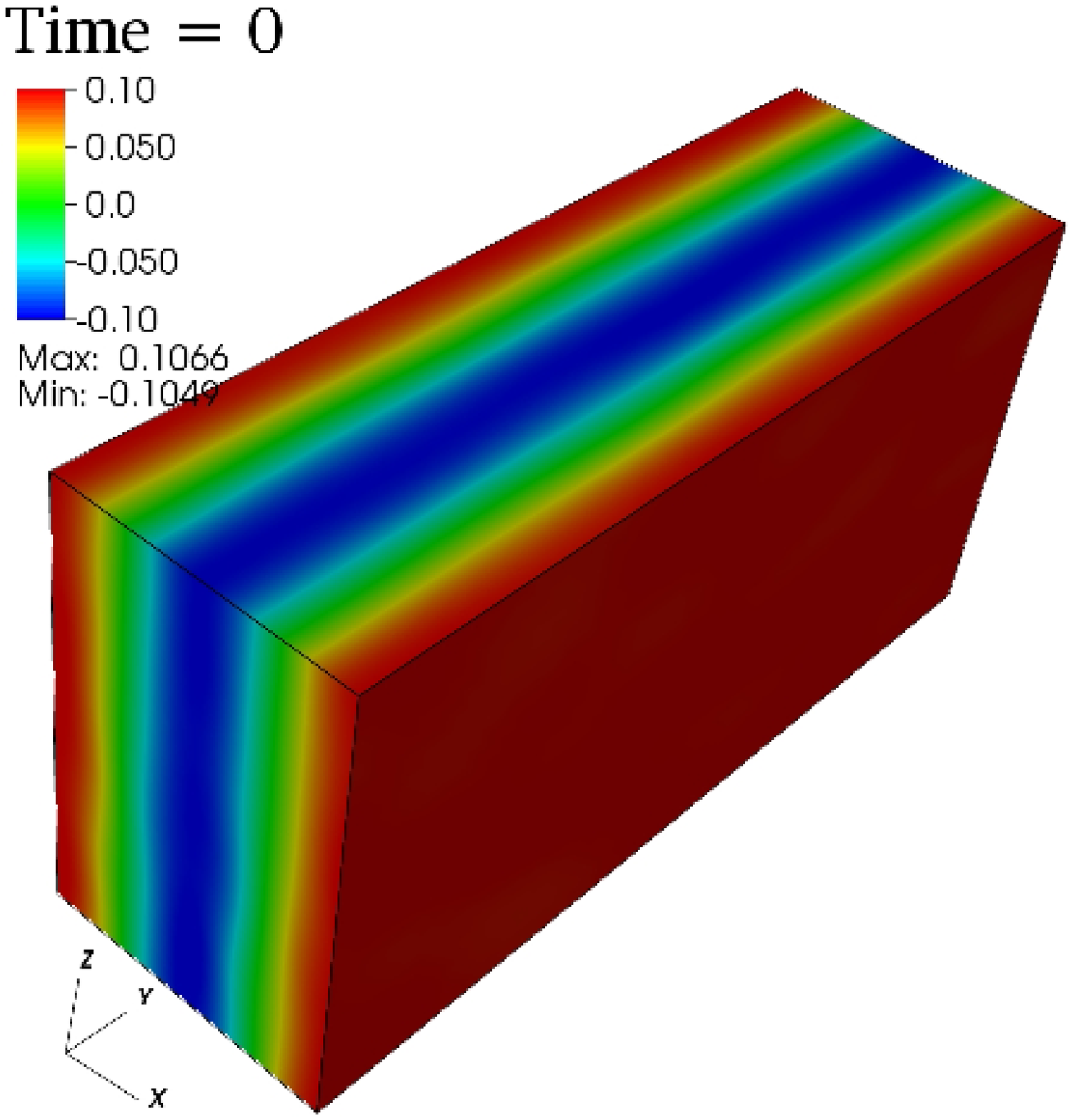}
  \includegraphics[width=.33\textwidth]{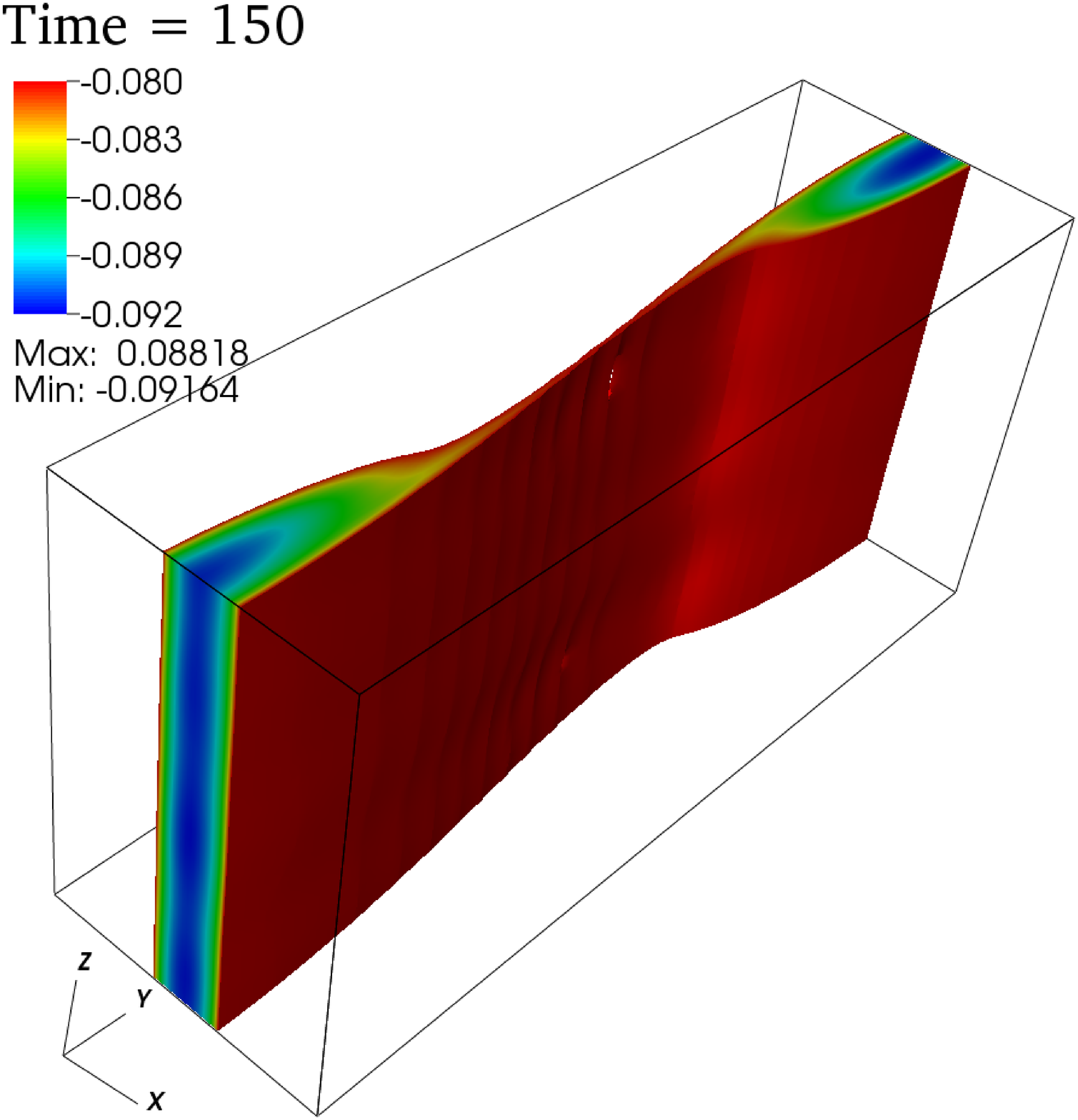}
  \includegraphics[width=.33\textwidth]{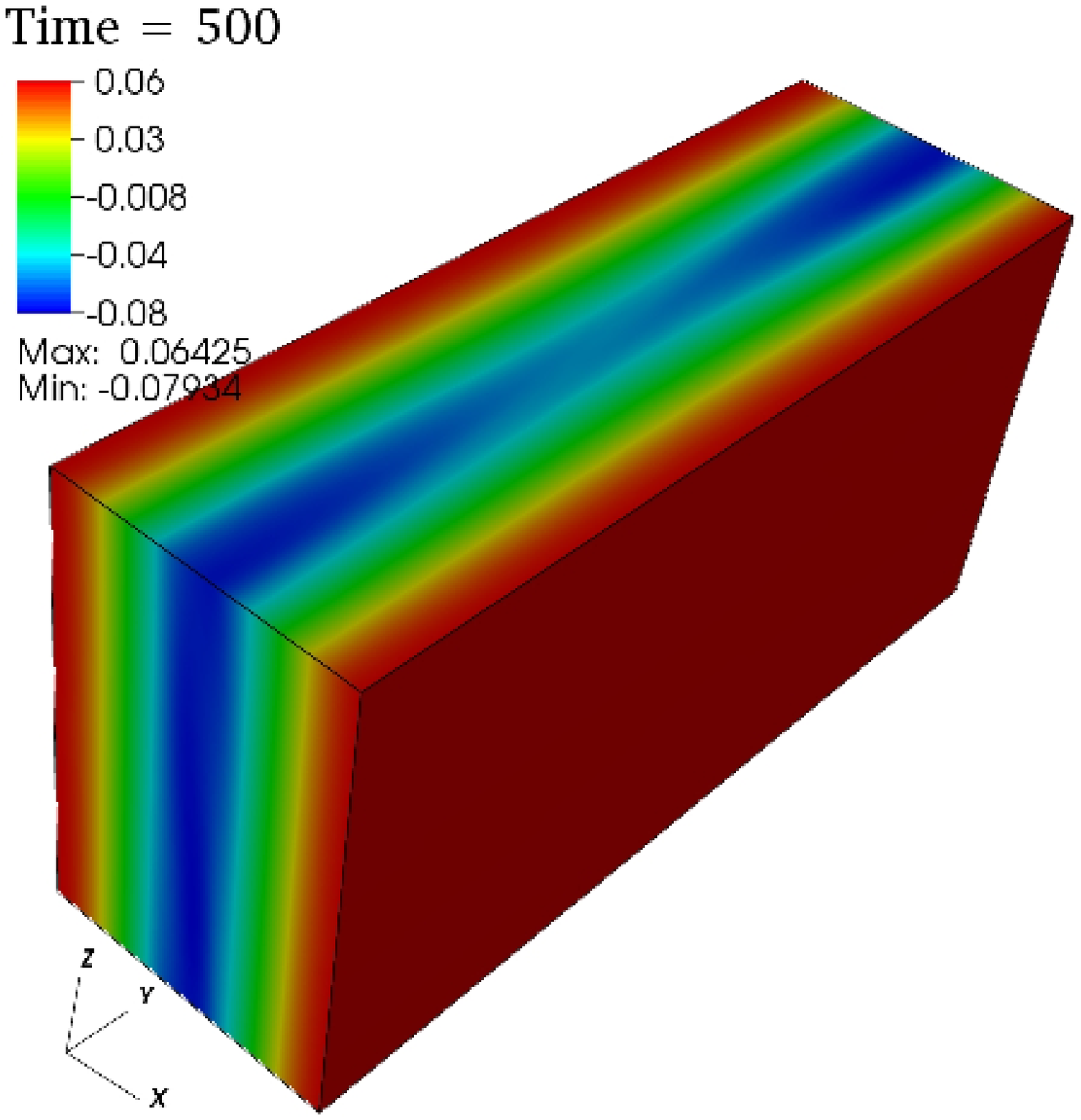}  
 \includegraphics[width=.32\textwidth]{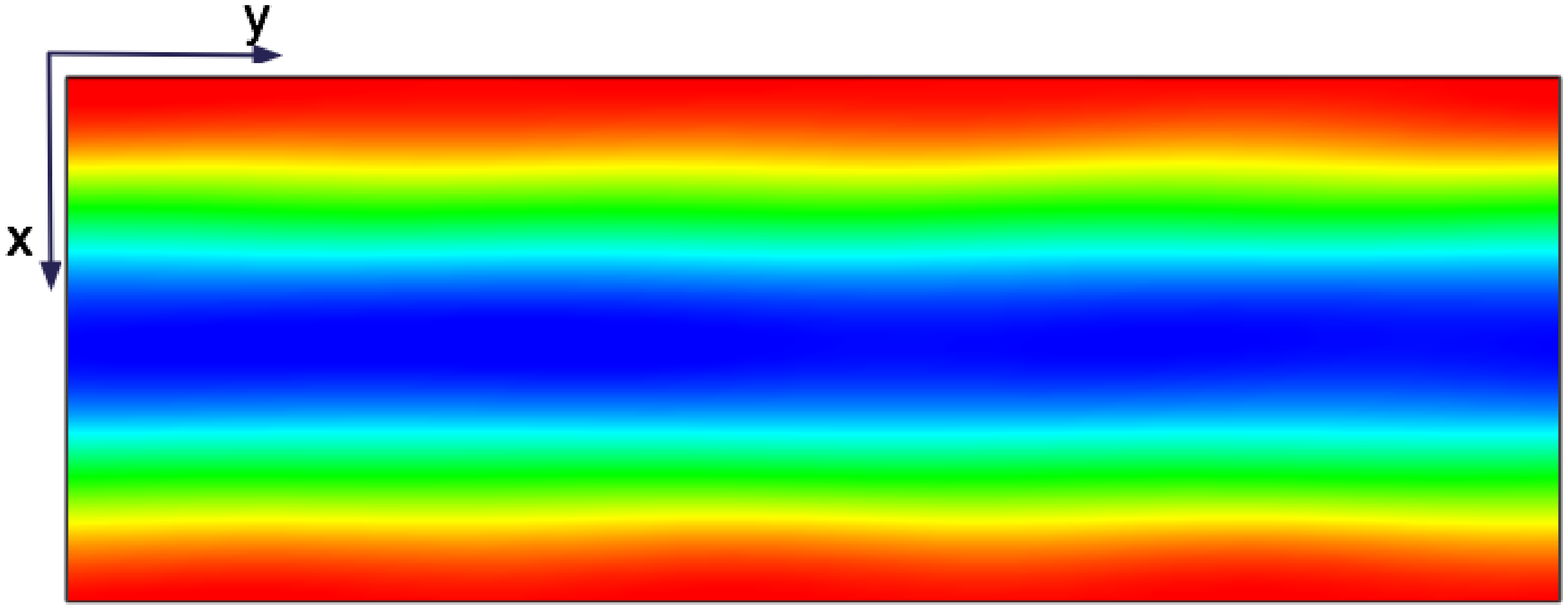}
  \includegraphics[width=.32\textwidth]{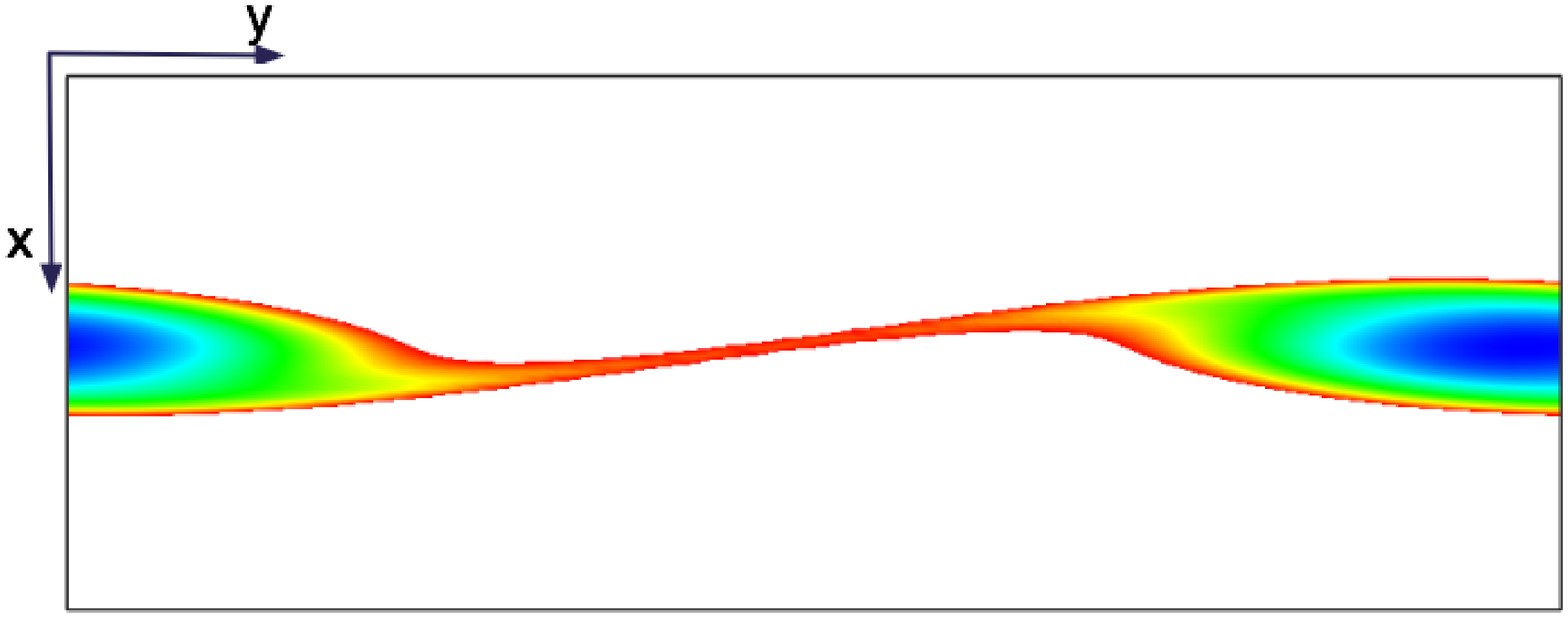}  
  \includegraphics[width=.33\textwidth]{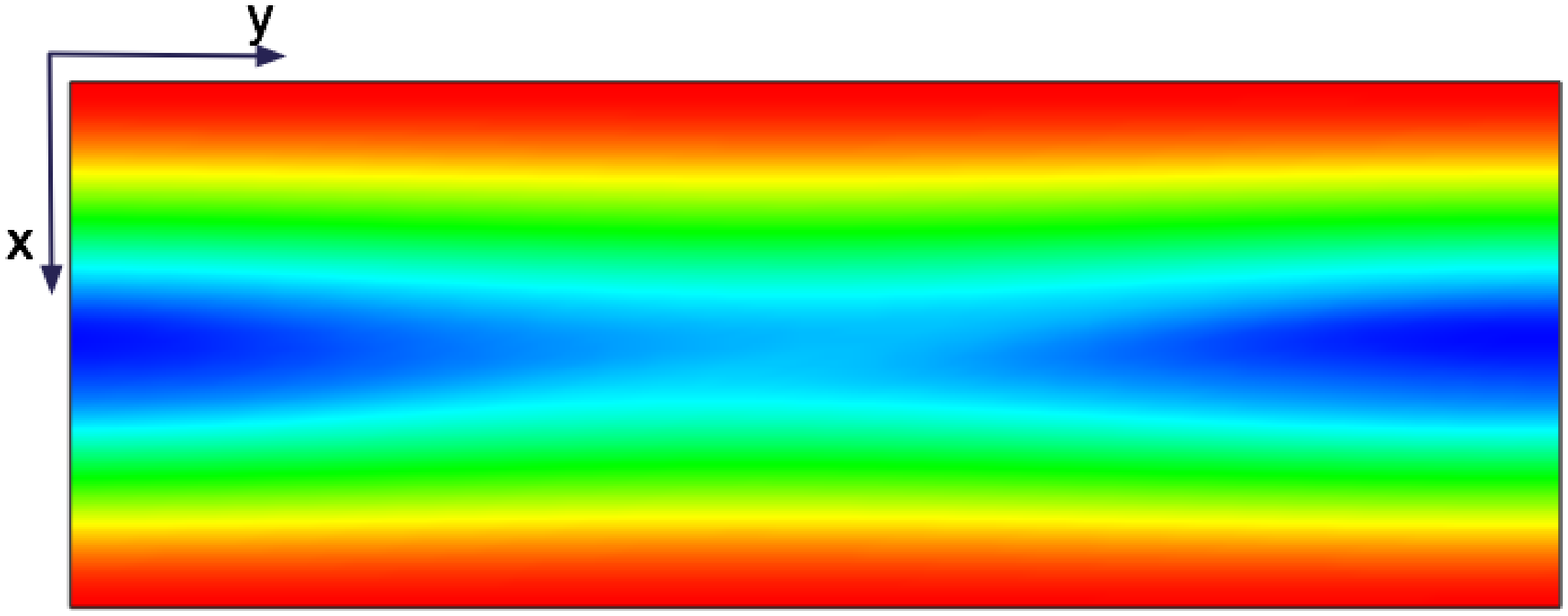}  
  \caption{{\bf Vortex Formation and Survival in a Short Box:} 
   Color depicts  $\omega_z$.
  The initial state is unstable to vertically symmetric ($k_z=0$) perturbations,
  and forms into a vortex.  But it is stable to 3D ($k_z\ne 0$) perturbations, and the evolution remains
  two-dimensional.  The bottom panels show  horizontal slices through the boxes in the upper
  panels, midway through the boxes.
   At time=150, the vortex has already
  formed. Only fluid with $\omega_z<-0.08$ is shown  in the middle panels to
  highlight the vortex, and to illustrate that surfaces of constant
  $\omega_z$ remain purely vertical.  At time=500,
  the vortex still survives.  Its amplitude is slowly decaying by viscosity, which
  acts on timescale=1130.  We set  $\Omega=1$ and $q=3/2$.
   The number of modes in the simulation is
  $n_x\times n_y\times n_z=64\times 64\times 32$, and the size of the simulation box is $(L_x,L_y,L_z)=({1\over 15},1,{1\over 2})$. In this figure, $L_z$ is to scale relative to
  $L_y$, but $L_x$ has been expanded by a factor of 5 for clarity.
}
  \label{fig:shortsim}
\end{figure*}

\begin{figure*}
  \centering
  \includegraphics[width=.32\textwidth]{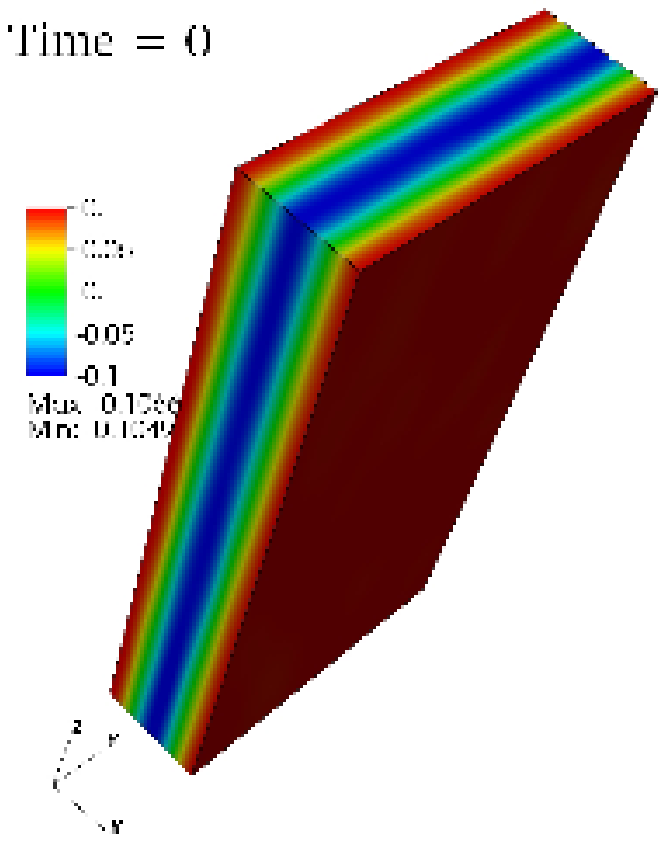}
  \includegraphics[width=.33\textwidth]{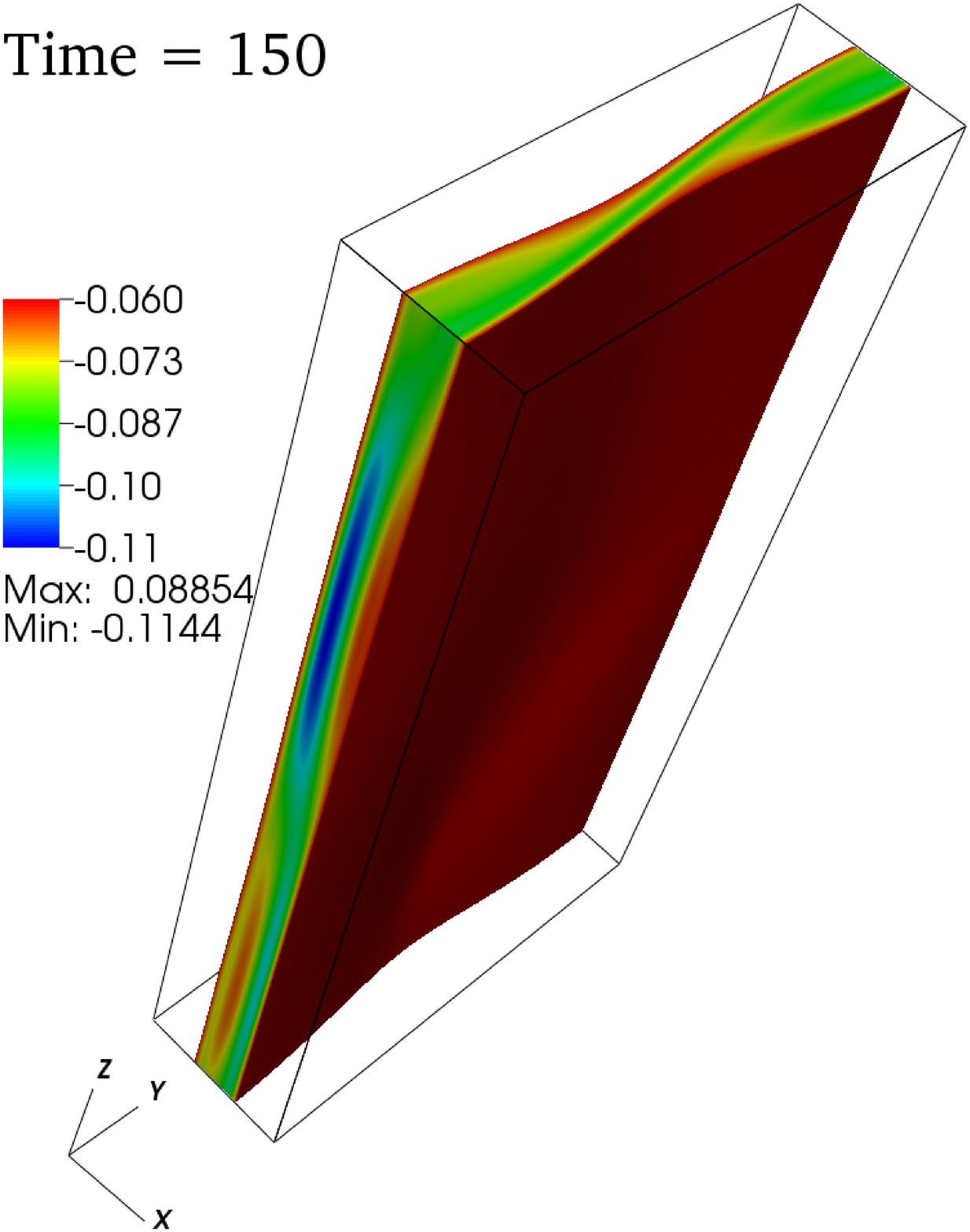}
  \includegraphics[width=.33\textwidth]{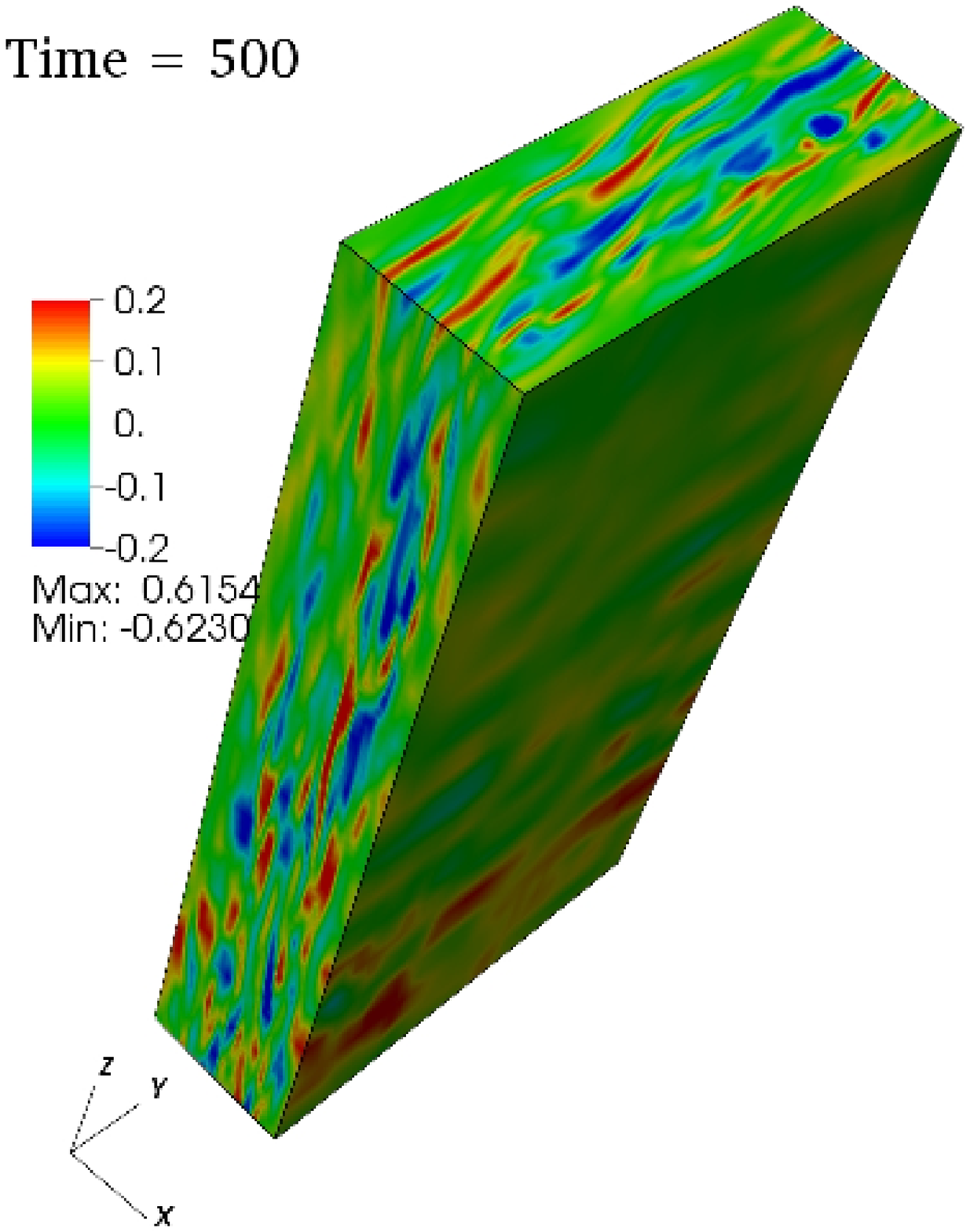}  
  \includegraphics[width=.32\textwidth]{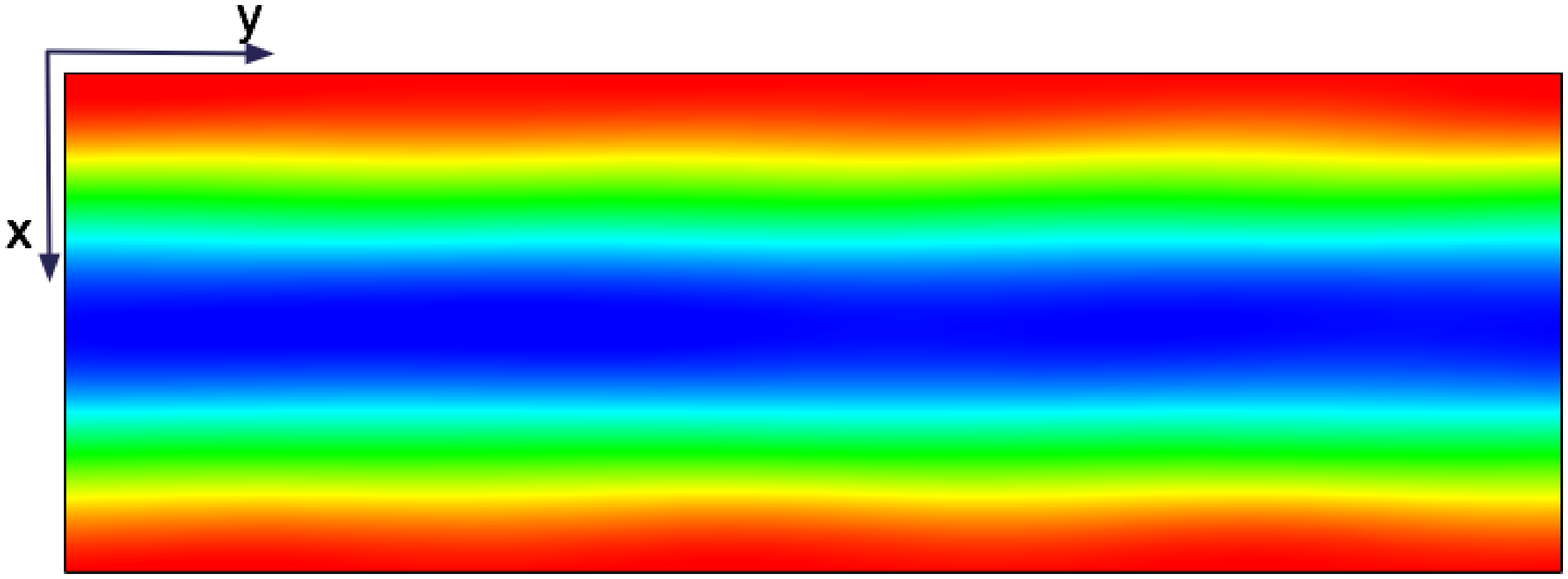}
  \includegraphics[width=.32\textwidth]{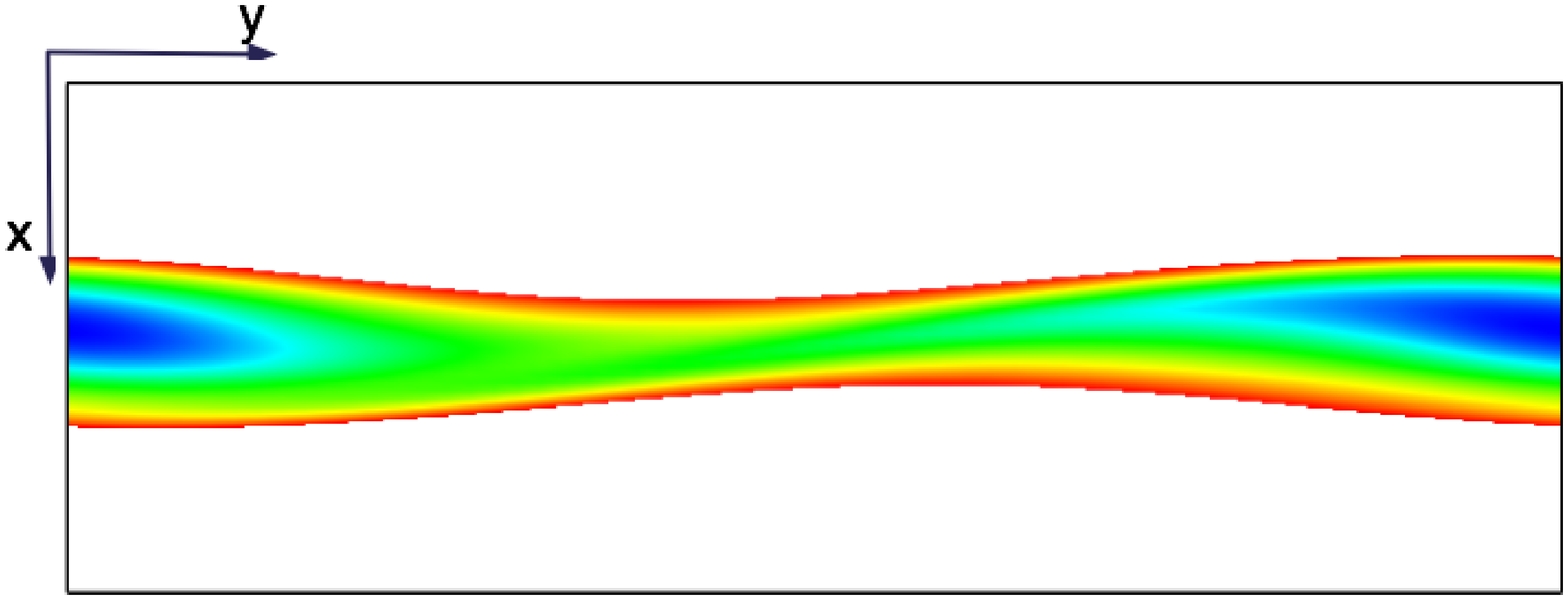}  
  \includegraphics[width=.33\textwidth]{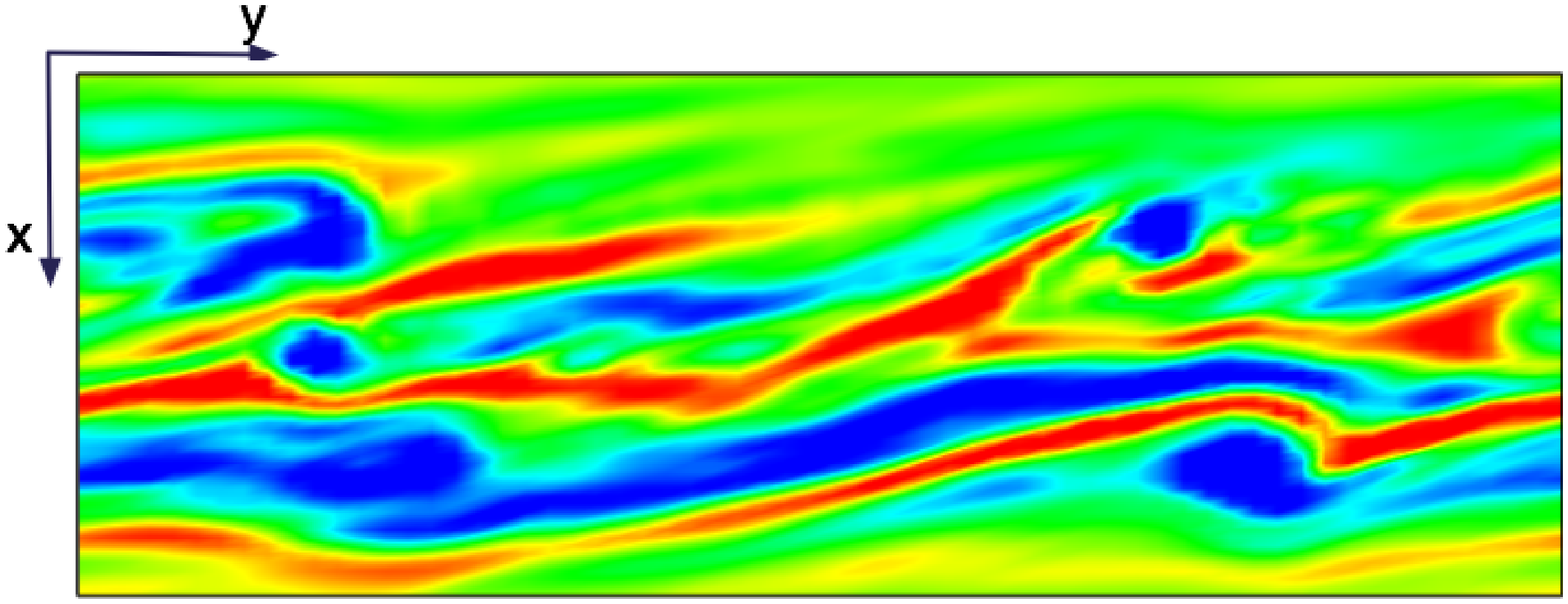}  
  \caption{{\bf Vortex Destruction in a Tall Box:} The setup 
  is identical to the short-box simulation of Figure \ref{fig:shortsim}, except that the height
  $L_z$ has been increased by a factor of 4, so that it now exceeds $L_y$. 
   The resulting evolution is dramatically different.
  The initial state is now unstable not only to 2D perturbations, but to 3D ones as well.  In the
  middle panels, surfaces of constant $\omega_z$ are warped, and the evolution is
  no longer vertically symmetric. 
    In the right panels, the flow looks turbulent.}
  \label{fig:tallsim}
\end{figure*}

\begin{figure}
  \hspace{-1cm}\includegraphics[width=.55\textwidth]{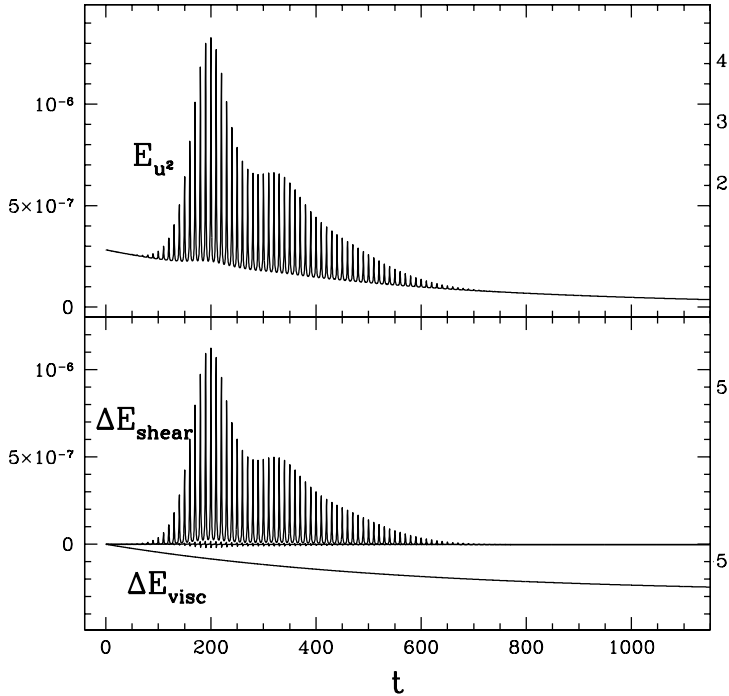}
  \caption{{\bf Energy in the Short Box:}
  The three contributions to the energy budget, $E_{u^2}$, $\Delta E_{\rm shear}$,
  and $\Delta E_{\rm visc}$, are defined in equations (\ref{eq:e1})-(\ref{eq:e3}).
  $E_{u^2}$ initially decays, and then rises to a peak
  near $t\sim 200$ as nonaxisymmetric perturbations turn the axisymmetric
  mode into a vortex. 
  Subsequently, the vortex decays due to viscosity.  The spikiness
  of the evolution is due to the boundary conditions, as explained in the text.
  Also shown in the bottom panel is the error due to numerical effects,
  $\Delta E_{\rm error}$ (eq. [\ref{eq:ee}]).
  It is unlabelled because it is mostly obscured by $\Delta E_{\rm shear}$.
  But it is nearly equal to zero everywhere, showing that the code accurately
  tracks the components of the energy budget.
  }
  \label{fig:shorten}
\end{figure}

\begin{figure}
  \hspace{-1cm}\includegraphics[width=.55\textwidth]{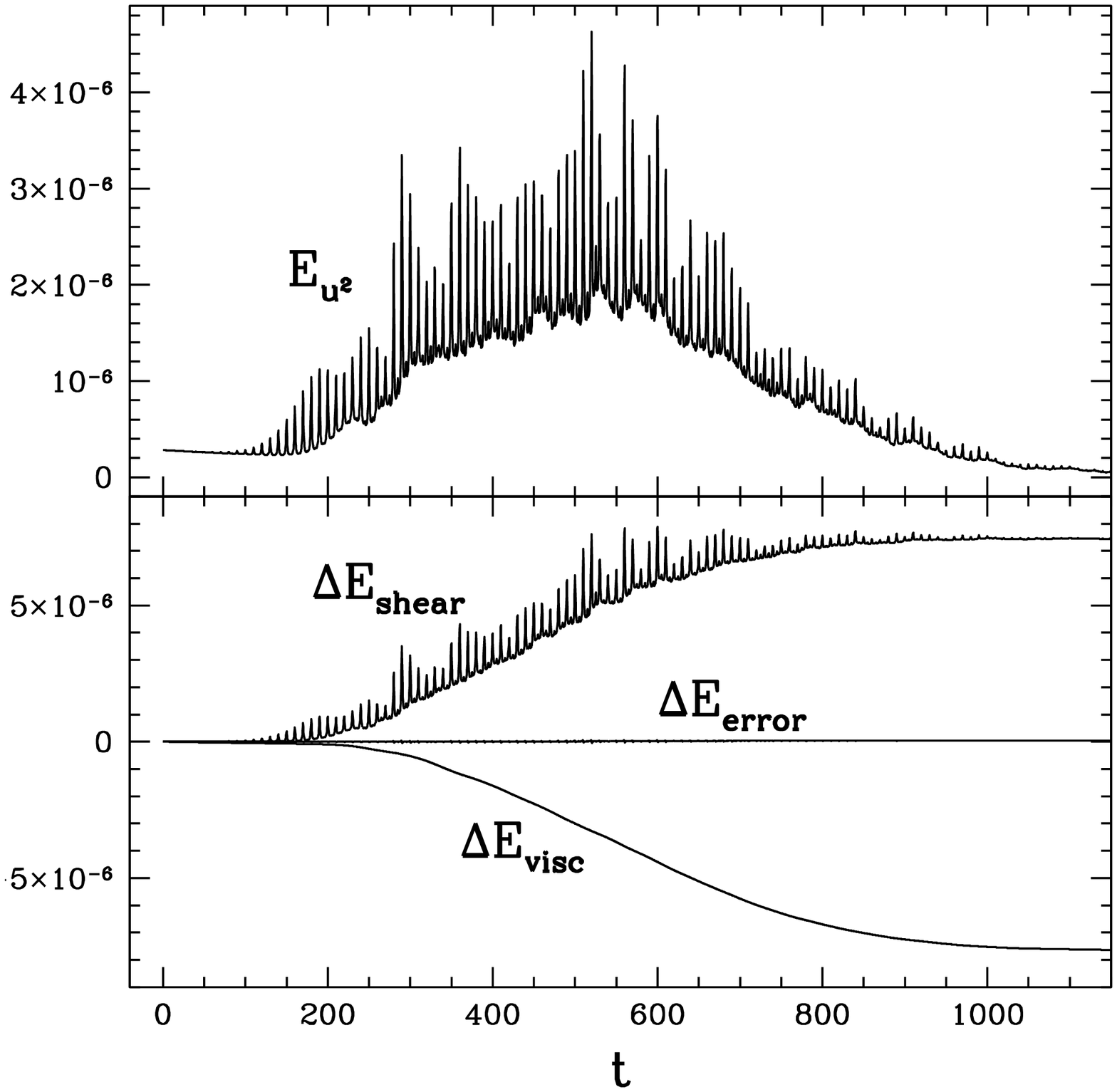}
  \caption{{\bf Energy in the Tall Box:} The initial evolution is almost the same
  as that seen in the short box (Fig. \ref{fig:shorten}).  But 3D perturbations are unstable and force
  the destruction of the vortex.   In the time interval $300\lesssim t\lesssim 600$,
  while the initial axisymmetric disturbance decays in a turbulent-like state,  the value of $E_{u^2}$ is significantly larger than its
  initial value, and $\Delta E_{\rm shear}$ rises nearly linearly in time, corresponding
  to nearly constant outwards transport of angular momentum in a disk.  
  The contribution of numerical errors to the time-integrated energy budget, $\Delta E_{\rm error}$ (eq. [\ref{eq:ee}]),
  remains small throughout.
  }
  \label{fig:tallen}
\end{figure}

\section{Two Pseudospectral Simulations}
\label{sec:pseudo}

The pseudospectral code is described in detail in the Appendix of L07.  
It solves the velocity-pressure equations of motion with an explicit viscous term
\be
\nu\nabla^2{\bld{u}}
\label{eq:visc}
\ee
added to equation (\ref{eq:eom2}).
(In L07, we did not include this term because we
only considered inviscid flows.)
The equations are solved 
in Fourier-space by decomposing fields into
 spatial Fourier modes whose wavevectors
are advected by the background flow $-qx\bld{\hat{y}}$.
As a result, the boundary conditions are periodic in the $y$ and $z$ dimension, and
``shearing periodic'' in $x$.
Most of our techniques are standard \citep[e.g.,][]{MG01,Rogallo81,BM06}.  One exception is 
our method for
remapping 
highly trailing wavevectors into highly leading ones, which is both simpler
and more accurate than the usual method.  In addition, our remapping 
does not introduce power into leading modes, because a mode's amplitude
has always been set to zero before the remap.
The code was extensively tested on 2D flows in L07.
A number of rather stringent 3D tests are performed in this  paper.  We shall show 
that the code correctly reproduces the linear evolution of 3D modes (\S \ref{sec:lin}), as
well as the nonlinear coupling between them (\S \ref{sec:nonlin}). We also show in 
the present section
that it tracks
the various contributions to the energy budget, and that the sum
of the contributions vanishes to high accuracy.

Figures \ref{fig:shortsim}-\ref{fig:tallen} show results 
  from two pseudospectral 
simulations.  One simulation illustrates the formation
and survival of a vortex, and the other illustrates
 vortex destruction.
In the first (the ``short box''),
 the number of Fourier modes  used 
is $n_x\times n_y\times n_z=64\times 64\times 32$, 
and the simulation box has dimensions
$L_x={1\over 15}$, $L_y=1$, and $L_z={1\over 2}$.  
In the second (the ``tall box''), the setup is identical,
except that it has $L_z=2$ instead of $1/2$.
Both simulations are initialized by setting
\be
\omega_z\Big\vert_{t=0}=-0.1\cos\left({2\pi\over L_x}x\right) \ .
\label{eq:initz}
\ee
In addition,  small perturbations are added to long-wavelength modes.
Specifically, labelling the wavevectors as
\be
(k_x,k_y,k_z)=2\pi\left({j_x\over L_x},{j_y\over L_y},{j_z\over L_z}  \right) \ ,
\label{eq:kj}
\ee 
with integers $(j_x,j_y,j_z)$,
we select all modes that satisfy
 $|j_x|\leq 3$, $|j_y|\leq 3$,
and $|j_z|\leq 3$, and set the Fourier amplitude of their $\omega_z$  to 
 $10^{-4}e^{i\phi}$, where $\phi$ is a random phase. 
But
we exclude the  $(j_x,j_y,j_z)=(0,0,0)$ mode, as well as 
$(j_x,j_y,j_z)=(\pm 1,0,0)$, which is  given by equation (\ref{eq:initz}).
Finally, we set $\Omega=1$, $q=3/2$, $\nu=10^{-7}$, and integration 
timestep $dt=1/30$.

With our chosen initial conditions, the mode given by equation (\ref{eq:initz})
is nonlinearly unstable to vertically symmetric ($k_z=0$) perturbations, and hence
it tends to wrap up into a vortex. From the approximate criterion for instability 
 (eq. [\ref{eq:2dinst}]),
we see that to illustrate the wrapping up into a vortex
of a mode with a small amplitude, one must make the simulation box
elongated in the $y$-direction relative to the $x$-scale of the mode in equation
(\ref{eq:initz}).

In the short box (Fig. \ref{fig:shortsim}), the evolution proceeds just as it would in two dimensions. 
The initial mode indeed wraps up into a vortex, and 
 the evolution remains vertically
symmetric throughout.  Once formed, the vortex can live for ever in the absence
of viscosity.
But in our simulation, there is a slow viscous decay.  The timescale for viscous decay
across
the width of the vortex is
$\sim 1/\nu k_x^2=1130$, taking $k_x=2\pi/L_x$.

In the tall box (Fig. \ref{fig:tallsim}), the evolution is dramatically different.  
In this case, the initial state is unstable not only to 2D  perturbations,
 but to 3D $(k_z\ne 0)$
ones as well. In the middle panel of that figure, we see 
that instead of forming a vertically symmetric vortex as in the short box,
 surfaces of constant $\omega_z$ are 
warped. 
By the third panel, the flow looks turbulent.

Figures \ref{fig:shorten}-\ref{fig:tallen} shows the evolution of the
 energy in these simulations.
Projecting ${\bld u}$
onto
the Navier-Stokes equation (eq. [\ref{eq:eom2}]  with viscosity included),
and spatially averaging, we arrive at the energy equation
\be
{d\over dt}
{\left\langle u^2
\right\rangle\over 2}=q
\langle u_xu_y\rangle 
+\nu\langle
{\bld u\cdot}\nabla^2\bld{u}
\rangle \ ,
\ee
after applying the shearing-box boundary conditions, 
where angled brackets denote a spatial average.
The time integral of this equation is
\be
E_{u^2}-E_{u^2}\vert_{t=0}=\Delta E_{\rm shear}+\Delta E_{\rm visc} \ , 
\label{eq:inte}
\ee
where
\beqn
E_{u^2}&\equiv& {\left\langle u^2
\right\rangle\over 2}
\label{eq:e1}
 \\
\Delta E_{\rm shear}&\equiv&
q\int_0^t \langle u_xu_y\rangle dt' \\
\Delta E_{\rm visc}&\equiv&
\nu\int_0^t
\langle
{\bld u\cdot}\nabla^2\bld{u}
\rangle dt' \ . \label{eq:e3}
\eeqn
The pseudospectral code records  each of these terms, and
Figure \ref{fig:shorten} shows the result in the short box 
simulation.  At very 
early times, $E_{u^2}$ decays from its initial value due to viscosity. 
At the same time, the small vertically symmetric perturbations are growing exponentially, 
and they start to give order unity perturbations by $t\sim 150$, by which time
 a vortex has been formed (Fig. \ref{fig:shortsim}).  
 As time evolves, $E_{u^2}$ gradually decays due to viscosity on the viscous
 timescale $=1130$.   The evolution  is very spiky.  
 We defer a discussion of this spikiness to the end of this section.
 
 Figure \ref{fig:tallen} shows the result in the tall box. The early evolution of 
 $E_{u^2}$ is similar to that seen in the short box.  Both start with the same $E_{u^2}$,
 and an initial period of viscous decay is interrupted by exponentially growing 
 perturbations. But in the tall box, not only are vertically symmetric modes
 growing, but modes with $k_z\ne 0$  are growing as well. By $t\sim 150$, there is a distorted
 vortex that subsequently decays into a turbulent-like state.  The energy $E_{u^2}$
 rises to a value significantly larger than its initial one, and it continues to rise
 until $t\sim 600$, when it starts to decay.  Throughout the time interval
 $300\lesssim t\lesssim 600$, $\Delta E_{\rm shear}$ rises nearly linearly in 
 time, showing that $\langle u_xu_y\rangle$ is positive and  nearly constant.
 
It is intriguing that  $\langle u_x u_y\rangle$ is positive for hundreds of orbits,
because
 it suggests that decaying vortices might transport
angular momentum outwards in disks and hence drive accretion.
Understanding the level of the turbulence,
 its lifetime, 
 and its nature are  topics for future work.
 Here we merely address the sign of $\langle u_xu_y\rangle$.
The quantity  $\langle u_xu_y\rangle$ is the 
flux of $y$-momentum in the $+x$-direction (per unit mass and spatially averaged).
It corresponds to the flux of angular momentum
in a disk.  A positive
$\langle u_xu_y\rangle$ implies an outwards flux of angular momentum, 
as is required to drive matter inwards in an accretion disk.
 (Even though the shearing box cannot distinguish
 inwards from outwards,
the sign of the angular momentum within a box depends on which side of the shearing
box one calls inwards.  Therefore, outwards transport of (positive) angular momentum
is well-defined in a shearing box.)
In the shearing box, 
any force that tends to diminish the background shear
flow $-qx\bld{\hat{y}}$ necessarily transports $y$-momentum in the $+x$-direction.
Hence the fact that $\langle u_xu_y \rangle >0$ in Figure \ref{fig:tallen} shows
that the turbulence exerts forces that resist the background shear, as one 
might
expect on physical grounds.
One can also understand why   $\langle u_xu_y \rangle >0$ from
energy considerations. 
Since  $\Delta E_{\rm visc}<0$, as may be seen explicitly by
an integration by parts, i.e. $\langle \bld{u\cdot}\nabla^2\bld{u} \rangle
=-\sum_{i,j}\langle(\partial_j u_i)^2\rangle$, equation (\ref{eq:inte})
may be rearranged to read
\be
\Delta E_{\rm shear} = \left\vert \Delta E_{\rm visc}\right\vert
+E_{u^2}-E_{u^2}\vert_{t=0} \ .
\label{eq:ee1}
\ee
If  the turbulence reaches a steady state---as it approximately does in 
Figure \ref{fig:tallen} during the time interval
$300\lesssim t\lesssim 600$---then the last two terms in the 
above equation are nearly constant, whereas $\left\vert \Delta E_{\rm visc}\right\vert$
increases linearly with time.   Hence $\Delta E_{\rm shear}$ must also increase.
The fact that energy dissipation implies outwards transport of angular momentum
is a general property of accretion disks \citep[e.g., ][]{LP}.   Since turbulence always
dissipates energy, it must also transport angular momentum outwards.  
However, this argument can be violated if  an external energy source drives
the turbulence,
in which case one would have to add this energy to the left-hand side of equation (\ref{eq:ee1}).
For example, the simulations of \cite{SB96} show that convective disks can transport angular momentum
inwards when an externally imposed heat source drives the convection.

Also shown in the bottom panels of Figures \ref{fig:shorten}-\ref{fig:tallen}
is the integrated energy error
\be
\Delta E_{\rm error}\equiv \Delta E_{\rm shear}+\Delta E_{\rm visc}+E_{u^2}\vert_{t=0}-E_{u^2}
\label{eq:ee}
\ee
 due to 
numerical effects, which is seen to be small. 
 (In Figure \ref{fig:shorten}, $\Delta E_{\rm error}$ is not labelled because
the curve is mostly obscured by $\Delta E_{\rm shear}$; it can be seen near $t\sim 200$,
and is everywhere very nearly equal to zero.)
The fact that $\Delta E_{\rm error}$ nearly vanishes throughout the simulations is 
not guaranteed by the pseudospectral algorithm.  Rather, we have chosen
$\nu$ to be large enough that the algorithm introduces negligible  error into the
energy budget.  To be more precise, at each timestep in the pseudospectral code,
modes that have $|j_x|>n_x/3$ or $|j_y|>n_y/3$ or $|j_z|>n_z/3$, where $j_{x,y,z}$
are defined via equation (\ref{eq:kj}), have their amplitudes set to zero (``dealiased'').
This introduces an error that is analogous to
grid error in grid-based codes.  By choosing  $\nu$ to be sufficiently large, it
is the explicit viscosity that forces modes
with large $k$ to have small amplitudes, in which case
the dealiasing procedure has little effect on the dynamics.
Increasing the resolution $n_x\times n_y\times n_z$
 would allow a smaller $\nu$ to be chosen---implying
a larger effective Reynolds number---while keeping the energy error small.

The curves of $E_{u^2}$ show
 sharp narrow spikes every time interval $\Delta t=10$, with width $\sim 1$.
 Similar spikes have been seen in other simulations \citep{UR04,SSG06}, but
 they are stronger and narrower in our simulations because our simulation box
 is elongated.  These spikes are due to the
 shearing-periodic boundary conditions.  It is perhaps simplest to understand
 them by following the evolution in $k$-space, as we shall do in 
  \S \ref{sec:nonlin} (see also L07).  But for now, we explain their origin
   in real-space.
By the  nature of shearing-periodic boundary conditions,   associated with the simulation 
 box centered  at $x=0$ are  ``imaginary boxes'' centered
    at $x=jL_x$ with
 integer $j=\pm 1,\pm 2,\cdots$.  These imaginary boxes completely tile the $x-y$ plane,
 and each
  contains a virtual copy of the conditions inside the simulation box.
 The boxes move relative to the simulation box in 
 the $y$-direction, with the speed of the mean shear at the center of each box,
 $-qjL_x$.
 Therefore, in  the  time interval 
 $\Delta t = L_y/(qL_x)=10$, all the boxes line up.  When this happens, 
 the velocity field $\bld{u}$ that is induced by the vorticity within all the boxes 
 (via eq. [\ref{eq:ominv}]) becomes large, because all the boxes reinforce each other,
 and therefore $E_{u^2}$ exhibits a spike.
Even though the shearing-periodic boundary conditions that we use are 
somewhat artificial, 
we are confident that using more realistic open boundary conditions would not
affect the main results of this paper---and particularly not the stability of
 axisymmetric modes to 3D perturbations.
In L07, where we considered 2D dynamics, 
we investigated both open and shearing-periodic boundary conditions, and
showed explicitly that both give similar results.
We also feel that the boundary conditions likely do not affect the level
and persistence of the ``turbulence'' seen in Figure \ref{fig:tallen}.  However, 
this is less certain.  Future investigations should more carefully address the role
of boundary conditions.

\section{Linear Evolution}
\label{sec:lin}

\begin{figure*}
  \centering
  \includegraphics[width=1\textwidth]{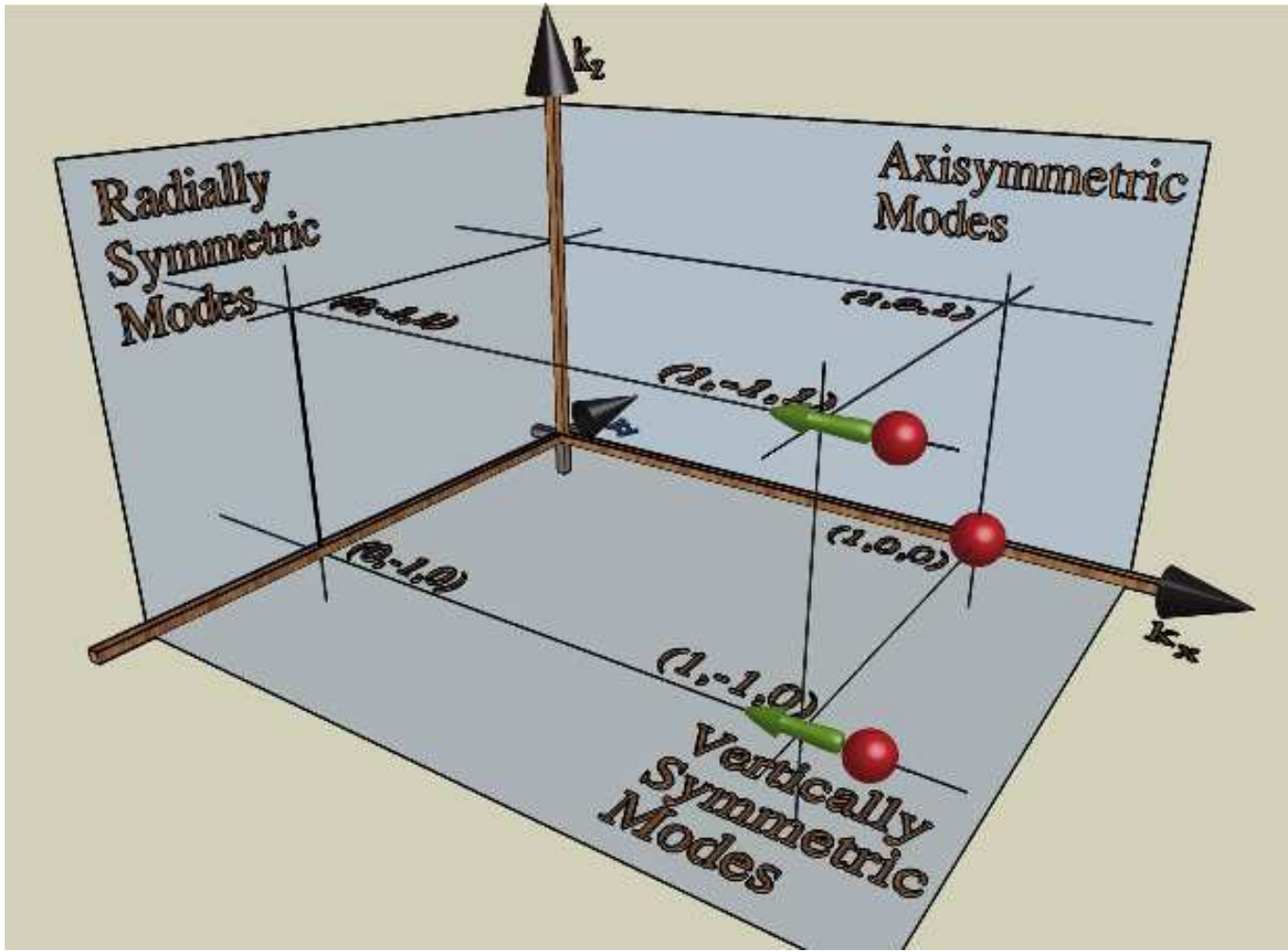}
  \caption{{\bf Evolution of Wavevectors:} Modes have constant $k_y$ and $k_z$, 
  and $k_x=qtk_y+$const.  The three spheres depict modes that
   play important roles in nonlinear
  instability.  The mode at $(1,0,0)$ does not move in $k$-space.  The other two modes
  are swinging modes that are depicted in the leading phase of their swing. They will
  become trailing after crossing through the radially symmetric plane. The mode
  crossing through $(1,-1,0)$ is responsible for 2D instability that forms vortices.
  The one crossing through $(1,-1,1)$ is responsible for 3D instability that destroys
  vortices.}
  \label{fig:kspace}
\end{figure*}

In the remainder of this paper, we develop a theory explaining the
stability of vortices seen in the above numerical simulations.  
We first consider the linear evolution of individual modes, 
and then proceed to show how nonlinear coupling between
linear modes can explain vortex stability.

The linear evolution has been considered previously \citep{AMN05,JG05,BH06}.
Only two aspects of our treatment are new.
First, we give the solution in terms
of variables that allow the simple reconstruction of
the full vectors $\bld{\omega}$ and $\bld{u}$.
And second, we give the analytic expression 
  for matching a leading mode 
onto 
a trailing mode that is
valid for all $k_y$ and $k_z$,

The linearized equation of motion is (eq. [\ref{eq:eomo}])
\be
(\partial_t-qx\partial_y)\bld{\omega}=-q\bld{\hat{y}}\omega_x+(2\Omega-q)\partial_z\bld{u} \ .
\label{eq:elin}
\ee
 A single mode may be written as
 \be
\bld{\omega}(\bld{x},t)=\bld{\hat{\omega}}(\bld{k_0},t)e^{i 
\left[
{\bld{k}}
\left(
{\bld{k_0}},t
\right)
\right]
{\bld\cdot}
\bld{x}
} \ , \label{eq:omk}
\ee
where $\bld{k_0}$ is a constant vector that denotes the wavevector at time $t=0$, and
the  wavevector $\bld{k}=\bld{k}(\bld{k_0},t)$ has components
\beqn
k_y &=& k_{0y}={\rm const} \\
k_z &=& k_{0z} ={\rm const}\\
k_x &=& {k_{0x}}+qtk_y\ne {\rm const} \ ,
\eeqn
so that upon insertion into equation (\ref{eq:elin}), the time-derivative of the exponential
cancels the term $-qx\partial_y{\bld{\omega}}$.
The velocity field induced by such a mode 
 is (eq. [\ref{eq:ominv}])
\beqn
\bld{u}(\bld{x},t)=
\bld{\hat{u}}(\bld{k_0},t)
e^{i \bld{k}\left[
\left(
{\bld{k_0}},t
\right)
\right]
{\bld\cdot}
\bld{x}} 
\label{eq:uk}
\eeqn
where
\be
\bld{\hat{u}}=
i{\bld{k\times\hat{\omega}}\over k^2}
\label{eq:udef}
\ee

Figure \ref{fig:kspace} sketches the evolution of wavevectors. 
Axisymmetric modes ($k_y=0$) do not move in $k$-space, as depicted by
the sphere at $(1,0,0)$ in Figure \ref{fig:kspace}.
 ``Swinging modes'' have $k_y\ne 0$, and their $k_x$ is time-dependent.  Their
fronts of constant phase are advected by the background shear.  Swinging modes
with $k_x/k_y<0$, as depicted by the two spheres 
near $(1,-1,0)$ and $(1,-1,1)$ in Figure \ref{fig:kspace},
 have phasefronts tilted into the background shear, i.e., they are leading modes.
   As time evolves, 
the shear first swings their $k_x$ through $k_x=0$, at which point their phasefronts
are radially symmetric.
Subsequently, they become
trailing modes ($k_x/k_y>0$), and approach alignment with the azimuthal direction
 ($k_x/k_y\rightarrow \infty$).

We turn now to the evolution of the Fourier amplitudes.
In the remainder of this paper we drop the hats
\beqn
\bld{\hat\omega}\rightarrow \bld{\omega} \ ,
\ \ 
\bld{\hat u}\rightarrow \bld{u} \ .
\eeqn
To distinguish real-space fields, we shall explicitly write their
spatial dependence, e.g. $\bld{\omega}(\bld{x})$.

Because $\bld{\omega}(\bld{x})$ is divergenceless, $\bld{{\omega}}$
only has two degrees of freedom, which we select to be
 ${\omega}_x$ and
\beqn
{\omega}_{yz}\equiv 
{\bld{\hat{x}\cdot}
\left({{\bld{k\times{\omega}}}}\right)\over k_{yz}} = 
\left\{
 \begin{array}
 {r@{\quad,\quad}l}
 -{\omega}_y & {\rm if\ } k_y=0 \\
{\omega}_z & {\rm if\ } k_z=0
 \end{array}
 \right.
\eeqn
where
\be
k_{yz}\equiv \sqrt{k_y^2+k_z^2} 
\ee
Our variable $\omega_{yz}$ is proportional to the variable $U$ of
\cite{BH06}. Adopting $\omega_x$ as the second degree of freedom
enables
  the full vectors to 
be reconstructed  as
\beqn
\bld{{\omega}} &=& 
-{\omega}_x
{\bld{k\times}(\bld{k\times \hat{x}})\over k_{yz}^2}
-{{\omega}_{yz}}{\bld{k\times\hat{x}}\over k_{yz}}
\label{eq:omrec}
\\
\bld{{u}} &=& -i\omega_{yz}{\bld{k\times}(\bld{k\times \hat{x}})\over k^2 k_{yz}}+
i\omega_x {\bld{k\times\hat{x}}\over k_{yz}^2}
\label{eq:urec}
\eeqn

The linearized equation  (\ref{eq:elin}) is expressed in terms of these degrees of freedom as
\be
{k_{yz}\over qk_y}{d\over dt}
\left(\begin{array}{c}{\omega}_x \\ {\omega}_{yz}\end{array}\right)
=\beta{\Omega\over \kappa}
\left(\begin{array}{cc}0 & -{1\over 2}{\kappa^2\over\Omega^2}{1\over 1+\tau^2} \\ 2 & 0\end{array}\right)
\left(\begin{array}{c}{\omega}_x \\ {\omega}_{yz}\end{array}\right) \ ,
\label{eq:linlim}
\ee
after introducing the  epicyclic frequency,
\beqn
\kappa \equiv \sqrt{2\Omega(2\Omega-q)}  \ ,
\eeqn
with $\kappa=\Omega$ in a Keplerian disks, and
\beqn
\tau&\equiv& {k_x\over k_{yz}} \\
\beta&\equiv&{\kappa\over q}{k_z\over k_y} \ .
\eeqn
As long as $k_y\ne 0$,
$\tau$ varies in time through its dependence on
$k_x=k_{0x}+qtk_y$.

For axisymmetric modes ($k_y=0$), $\tau$ is constant and
\be
{d^2\over dt^2}\omega_{yz}+\kappa^2{k_z^2\over k_x^2+k_z^2}\omega_{yz}=0 \ ,
\label{eq:axi3d}
\ee
the solution of which is sinusoidal with frequency $\kappa k_z/\sqrt{k_x^2+k_z^2}$.
Axisymmetric modes with phasefronts aligned with the plane of the disk ($k_x=k_y=0$) have
in-plane fluid velocities, and they oscillate at
the epicyclic frequency of a free test particle, $\kappa$.
  But axisymmetric modes with tilted
phasefronts have slower frequencies, because fluid pressure causes deviations from
free epicycles.  In the limit of vertical axisymmetric phasefronts ($k_z=k_y=0$), the effects of
rotation disappear entirely, and this zero-frequency mode merely alters the mean shear
flow's velocity profile.

\begin{figure} 
 \hspace{-1.cm} \includegraphics[width=.55\textwidth]{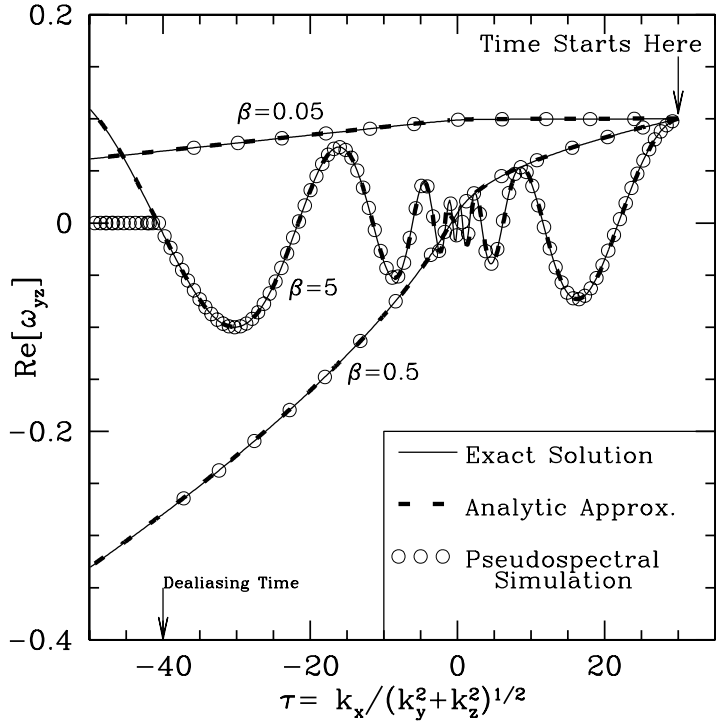}
  \caption{{\bf Linear Evolution of Mode Amplitudes for Three Values 
  of $\beta$:} Time runs from right to left.
  Solid curves show the exact, numerically integrated solution of equation (\ref{eq:bh}).
  The initial value of $d\omega_{yz}/d\tau$ was chosen so that $\omega_B=0$ initially
  (eq. [\ref{eq:powerlaw}]). 
    Dashed lines show the analytic solution (eq. [\ref{eq:powerlaw}])
  with constant $\omega_A$ and $\omega_B=0$ for $\tau>0$; while for $\tau<0$, the
  normal mode amplitudes are set to
  different constants that are given by equation (\ref{eq:trans}). 
  We exclude the domain $|\tau|<1$ from the dashed curve, because the analytic
  approximation does not apply there.
   Circles show output
  from the pseudospectral code, integrated with a timestep $dt=1/15$ and with the viscosity
  set to zero. 
    }
  \label{fig:lin}
\end{figure}

For swinging modes
($k_y\ne 0$),  it is convenient to employ $\tau$ as  the time variable.
Since 
\be
{k_{yz}\over qk_y}{d\over dt}={d\over d\tau} \ ,
\ee
we have
\be
{d^2\over d\tau^2}\omega_{yz}+{\beta^2\over 1+\tau^2}\omega_{yz}=0 \ .
\label{eq:bh}
\ee
\citep{BH06}.
Figure \ref{fig:lin} plots numerical solutions of this equation, and shows that it matches 
the output from the pseudospectral code, as well as the analytic theory described below.
Given $\omega_{yz}$, it is trivial to construct $\bld{\omega}$ and $\bld{{u}}$
from
\be
\omega_x={\kappa\over 2\beta\Omega}{d\omega_{yz}\over d\tau}
\label{eq:oxeq}
\ee
and equations (\ref{eq:omrec}) and (\ref{eq:urec}).

For highly leading or trailing modes ($|\tau|\gg 1$), equation (\ref{eq:bh}) has simple power-law solutions,
\beqn
\omega_{yz}&=&\omega_A|\tau|^{1-\delta\over 2}+\omega_B|\tau|^{1+\delta\over 2} \ , 
\ \ |\tau|\gg 1 \label{eq:powerlaw}
\eeqn
\citep{BH06}, where  $\omega_A$ and $\omega_B$ are constants that
we shall call the ``normal-mode'' amplitudes, and
\be
\delta \equiv \sqrt{1-4\beta^2} \ ,
\ee
which is imaginary for $|\beta|>1/2$.
As a mode's wavevector evolves along a line in $k$-space, its
amplitude is oscillatory if this line is much closer to the $k_z$ axis
than to the $k_y$ one, and  non-oscillatory if the converse is true.
The transition occurs at $|\beta|=1/2$.     This
behavior may be understood as a competition between shear and epicyclic oscillations.  
The timescale for $k_x$ to change by an order-unity factor due to the
shear is $t_{\rm shear}\sim |k_x/\dot{k}_x|=|k_x/qk_y|$, and
the timescale for epicyclic oscillations of axisymmetric modes
 is $t_{\rm epi}\sim \kappa^{-1}|k_x/k_z|$ for 
$|k_x|\gg |k_z|$.  
Therefore $|\beta|\sim t_{\rm shear}/t_{\rm epi}$, and
when $|\beta|\gg 1$ the epicyclic time is shorter and
so the mode's amplitude oscillates as its wavevector is slowly advected by the shear.  But
 when $|\beta|\ll 1$ the shear changes the wavevector faster
than the amplitude can oscillate.

The solution (\ref{eq:powerlaw}) breaks down in mid-swing.  
As a swinging wave changes from leading to trailing, its ``normal-mode
amplitudes''  change abruptly  on the timescale that $\tau$ changes from
$\pm 1$ to $\mp 1$ via
\be
\left(\begin{array}{c} \omega_A \\ \omega_B\end{array}\right)_{\rm trail}
=
\left(\begin{array}{cc}T_{AA} & T_{AB} \\T_{BA} & T_{BB}\end{array}\right)
\left(\begin{array}{c} \omega_A \\ \omega_B\end{array}\right)_{\rm lead}
\label{eq:trans}
\ee
where the transition matrix has components
\beqn
T_{AA}&=&-T_{BB}=\csc\left({\delta\pi/ 2}\right) \\
T_{BA}&=&
-{\cot(\delta\pi/2)^2\over T_{AB}}
=-2^{\delta+1}{1\over\delta}{1-\delta\over 1+\delta}
{\Gamma(1+\delta/2)^2\over \Gamma(1/2+\delta/2)^2}
\eeqn
and determinant $=-1$, and hence is its own inverse.
 The components are complex when $|\beta|>1/2$.
To derive these components, we took advantage of the fact that equation (\ref{eq:bh}) has
hypergeometric solutions   \citep{JG05,BH06}, and matched
these onto the normal-mode solution given above.  We omit
the unenlightening details.

\section{Nonlinear Evolution: Formation and Destruction of Vortices}
\label{sec:nonlin}

\subsection{Qualitative Description}

The instability that destroys  vortices is a generalization of 
the one that forms them. 
 We review here how vortices form,
before describing the instability that destroys them.  In \S \ref{subsec:stab},
we make this description quantitative.

Vortices form out of a nonlinear instability that involves vertically symmetric ($k_z=0$)
modes. (See L07 for more details of the 2D dynamics than are presented here.) 
  Consider the two vertically symmetric modes shown in Figure \ref{fig:kspace}: the
``mother'' mode at $(1,0,0)$ and the ``father'' mode that is depicted crossing through $(1,-1,0)$.
Triplets of integers 
$(j_x,j_y,j_z)$
label  values of wavevectors $(k_x,k_y,k_z)$
(for example, via [\ref{eq:kj}]).
The mother is both axi- and vertically-symmetric, and the father is a leading swinging mode.

As the father swings through radial symmetry, i.e. as it crosses through
the point $(0,-1,0)$, its
 velocity
field  is
strongly amplified by the background shear.  This can be seen from  
 \S \ref{sec:lin}, which shows that swinging modes with $k_z=0$ have
 $\omega_{yz}$=const., and hence $u_x=i(\omega_{yz}/k_{yz})/(1+\tau^2)$, which
 becomes largest when $\tau$ crosses through 0.
When the father is near the peak of its transient amplification ($|\tau|\lesssim 1$), 
it couples most strongly with the mother, and they produce a ``son''
near $(1,-1,0)=(1,0,0)+(0,-1,0)$.  The son will then swing through radial symmetry
where it will  couple (oedipally) with the mother to produce a grandson near $(1,-1,0)$,
which can repeat the cycle.
We summarize this 2D instability feedback loop  as
\beqn
{\rm linear\ amplification:}& (1,-1,0)\rightarrow (0,-1,0)  \nonumber \\
{\rm nonlinear\ coupling:}& (0,-1,0)+(1,0,0)\rightarrow(1,-1,0) \nonumber
\eeqn
The criterion for instability is simply that the amplitude of the son's $\omega_{yz}$ be
larger than that of the father.  
As shown in L07, if instability is triggered, its nonlinear outcome 
in two dimensions is a long-lived vortex.

The three-dimensional instability that is responsible for destroying vortices
is a straightforward generalization.   
The mother mode is still at $(1,0,0)$, but now the father mode starts near 
$(1,-1,1)$.
Symbolically, the feedback loop is
\beqn
{\rm linear\ amplification:}& (1,-1,1)\rightarrow (0,-1,1)  \nonumber \\
{\rm nonlinear\ coupling:}& (0,-1,1)+(1,0,0)\rightarrow(1,-1,1) \nonumber
\eeqn
The 2D instability described above is just a special case of this 3D one in the limit that $k_z=0$.
In general, the stability of a mother  mode at $(1,0,0)$ with given $k_x=\bar{k}_x$ and 
${\bld\omega}=\bld{\bar{\omega}}$ 
depends on the $k_y$ and $k_z$ of the 
 father-mode perturbations (as well as on the parameters
$q$ and $\Omega$). Which $k_y$ and $k_z$ are accessible in turn depends
 on the dimensions $L_y\times L_z$ of the simulation box---or equivalently,
on the circumferential distance around a disk and the scale-height.
In \S \ref{subsec:stab}, we map out quantitatively  the  region in the
$k_y-k_z$ plane that leads to instability.
For now, it suffices to note that the unstable region has $|k_y|\lesssim |\bar{k}_x \bar{\omega}|/q$ and
$|k_z|\lesssim |k_y|$.  
We conclude that a given mother mode suffers one of three
possible fates, depending on $L_y$ and $L_z$.
\begin{enumerate}
\item
If $L_y$ is less than a critical value 
($\sim q/|\bar{\omega}\bar{k}_x|$), then the mother mode is
stable to all perturbations. 
\item
 If $L_y$ is larger than this critical value, then the 
mother mode is unstable to vertically symmetric ($k_z=0$) perturbations;  if in addition $L_z$ is sufficiently
small that all modes with $k_z\ne 0$ are stable, then the  mother mode turns
into a long-lived vortex (Figure \ref{fig:shortsim}).
 \item If both $L_y$ and $L_z$ are sufficiently large, the mother mode
is unstable both to vertically symmetric and to 3D perturbations.  When this happens, the mother
starts to form a vortex, but this vortex is 3D-unstable.  The result is turbulence (Figure \ref{fig:tallsim}).
\end{enumerate}
There is also a possibility that is intermediate between numbers 2 and 3: if the conditions
described in number 2 hold, the essentially 2D dynamics that results
can nonlinearly produce new mother modes that are unstable to 3D perturbations.
In this paper, we shall not consider this possibility further, since it did not occur in
the pseudospectral simulations of \S \ref{sec:pseudo}.  
We merely note that in our simulations
of this possibility (not presented in this paper), we found that when the new mother modes
decayed, they also destroyed the original mother mode.

\subsection{The Stability Criterion}
\label{subsec:stab}

\begin{figure}
  \hspace{-1cm}\includegraphics[width=.55\textwidth]{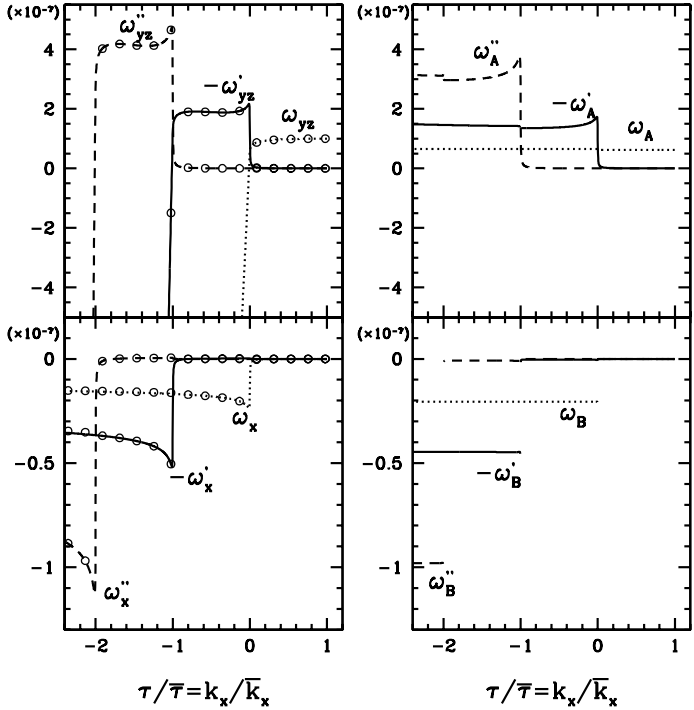}
  \caption{{\bf Nonlinear Evolution of 3D Instability:} Time runs from right to left.  In the two left panels, lines show numerical 
  solutions of equations (\ref{eq:lin2}) and (\ref{eq:soneq}), as well as the grandson's
  equation. Also shown as circles
  are  the output from a pseudospectral simulation, showing excellent agreement
  with the ``exact'' solutions.    The
  following parameters have been chosen: $\Omega=1$, $q=3/2$, $\bar{\omega}=0.005$, $\bar{k}_x=2\pi\cdot 15$,
  $k_y=-2\pi/30$, $k_z=0.45|k_y|\Rightarrow \beta=-0.3$. The small disagreement
  between pseudospectral and exact solutions  for $\omega_x'$ at
  $\tau<-2\bar{\tau}$ is due to the conjugate modes that, for simplicity,  we have not included in equations 
  (\ref{eq:lin2}) and (\ref{eq:soneq}); see footnote \ref{foot:cc}.   The two right panels show the mode amplitudes, defined via equation (\ref{eq:sonnm})
  for the son, and similarly for the father and grandson. Although these two right
  panels contain the same information as the left ones, they are helpful
  in constructing the analytic form of the growth factor $\chi$ (see Appendix).  With the parameters
  chosen for this figure, equation (\ref{eq:amp}) predicts $\chi=-2.2$ for the amplification
factor between successive generations, in agreement with that seen in the figure.
  }
  \label{fig:nonlin}
\end{figure}

To quantify the previous discussion, we choose an initial state as
in Figure \ref{fig:kspace}, with the mother
mode at $(1,0,0)$ and the father a leading mode crossing through $(1,-1,1)$. 
 The son mode, not depicted in the figure,
 is initially
crossing through the point $(2,-1,1)$.  We set its initial vorticity---as well as the
initial vorticity of all modes other than the mother and father---to 
zero.\footnote{\label{foot:cc}We
ignore the complex conjugate modes for simplicity.
Since 
$\bld{\omega(x)}$ is real-valued, each mode with wavevector 
and amplitude $(\bld{k},\bld{\omega)}$ is accompanied by a
conjugate mode that has $(-\bld{k},\bld{\omega}^*)$.
In our initial state, there are really four modes with non-zero amplitudes: the mother
at $(1,0,0)$ and its conjugate at $(-1,0,0)$, and the father and its conjugate.  We may ignore the conjugate
modes because they do not affect the instability described here.  
As shown in L07 for the 2D case, 
their main effect  is that when the son swings through $(0,-1,1)$, not only 
does it couple with the mother at $(1,0,0)$ to produce a grandson at $(1,-1,1)$, but
it also couples with the conjugate mother at $(-1,0,0)$ to partially kill its father,
 which is then at $(-1,-1,1)$ (bringing to mind the story of Oedipus).
But since the father is a trailing mode at this time, it no longer participates
in the instability.  Nonetheless, the conjugate modes do play a role in the nonlinear
outcome of the instability.
}
The father's wavevector and Fourier amplitude are labelled as in \S\ref{sec:lin}, and
the mother's and son's  are labelled with bars and primes:
\beqn
{\rm father:\ }&{\bld{{k}}}& \ , \
{\bld{\omega}}
 \nonumber \\
{\rm mother:\ }&{\bld{\bar{k}}}&\equiv \bar{k}{\bld{\hat{x}}}  \ \  , \ 
{\bld{\bar{\omega}}}\equiv{\bar{\omega}}{\bld{\hat{z}}} 
\nonumber \\
{\rm son:\ }&{\bld{k'}}& \equiv \bar{k}{\bld{\hat{x}}} +{\bld{k}} \  , \ 
{\bld{\omega'}}
\nonumber
\eeqn
Note that $\bar{k}$=const., and $\bar{\omega}_x=0$ because the vorticity
must be transverse to the wavevector.
We also set $\bar{\omega}_y=0$; otherwise $\bar{u}_z\ne 0$, which corresponds to
 a mean flow out the top of the box  and  in through the bottom.

At early times, the father mode swings through the point $(0,-1,1)$. 
Since the only other nonvanishing mode at this time is the mother, there 
are no mode couplings  that can nonlinearly change the father's
amplitude.
Therefore its amplitude
is governed by the linear equation (\ref{eq:linlim}), which we reproduce here as
\be
{d\over d\tau}
\left(\begin{array}{c}{\omega}_x \\ {\omega}_{yz}\end{array}\right)
=\beta{\Omega\over \kappa}
\left(\begin{array}{cc}0 & -{1\over 2}{\kappa^2\over\Omega^2}{1\over 1+\tau^2} \\ 2 & 0\end{array}\right)
\left(\begin{array}{c}{\omega}_x \\ {\omega}_{yz}\end{array}\right) \ .
\label{eq:lin2}
\ee
During its swing, it couples with the mother to change the amplitude of the son. 
 The linear part  of the son's evolution is given by the
above equation with primed vorticity and wavevector in place of unprimed.
The nonlinear part is given by \be
{d\over dt}{\bld{\omega'}}\Big\vert_{\rm nonlin}=i\bld{
k'\times(\bar{u}\times \omega+u\times\bar{\omega})
}
\ee
(eq. [\ref{eq:eomo}])
where  $\bld{\bar{u}}=-i(\bar{\omega}/\bar{k}){\bld{\hat{y}}}$ and
 $\bld{u}=i\bld{k\times\omega}/k^2$ (eq. [\ref{eq:udef}]).
Adding the linear and nonlinear parts, and re-expressing in terms of our chosen
degrees of freedom, we find
\beqn
{d\over d\tau}
\left(\begin{array}{c}{\omega}_x' \\ {\omega}_{yz}'\end{array}\right)
=\beta{\Omega\over \kappa}
\left(\begin{array}{cc}0
\ \ \ 
 & -{1\over 2}{\kappa^2\over\Omega^2}{1\over 1+(\tau+\bar{\tau})^2} \\ 
2\ \ \ 
  & 0\end{array}\right)
\left(\begin{array}{c}{\omega}_x' \\ {\omega}_{yz}'\end{array}\right) 
\nonumber
\\
-{\bar{\omega}\over q}
\left(\begin{array}{cc} {1\over \bar{\tau}}
\ \ \ 
&{\beta q\over\kappa}{1\over 1+\tau^2} \\
 0\ \ \  
 &{1\over\bar{\tau}}- {\bar{\tau}\over 1+\tau^2}\end{array}\right)
 \left(\begin{array}{c}{\omega}_x \\ {\omega}_{yz}\end{array}\right)  \ ,
 \label{eq:soneq}
\eeqn
where the dimensionless constant
\be
\bar{\tau}\equiv {\bar{k}_x\over k_{yz}} \ 
\ee
depends on both the mother's and father's wavevectors.
It is the father's $\tau\equiv k_x/k_{yz}$ that is being used as the time-coordinate
for evolving the son's amplitude.
The grandson's equation is the obvious extension: 
denoting
the grandson's amplitudes with double primes, one need only
make the following replacements
in equation (\ref{eq:soneq}):
 $\bld{\omega'}\rightarrow\bld{\omega''}$,
$\bld{\omega}\rightarrow\bld{\omega'}$, and
$\tau\rightarrow \tau+\bar{\tau}$.
Subsequent generations
evolve  analogously.

The father's equation  (\ref{eq:lin2}) is easily solved, as shown in \S \ref{sec:lin}. 
Inserting this solution into equation (\ref{eq:soneq}) produces a linear inhomogeneous
equation for the son's amplitude, and similarly for the grandson's.
Figure \ref{fig:nonlin} plots numerical solutions of these equations.  Also shown
 as circles are output from a pseudospectral simulation,
showing excellent agreement.

\begin{figure*}
  \centering
  \hspace{-3cm}\includegraphics[width=.4\textwidth]{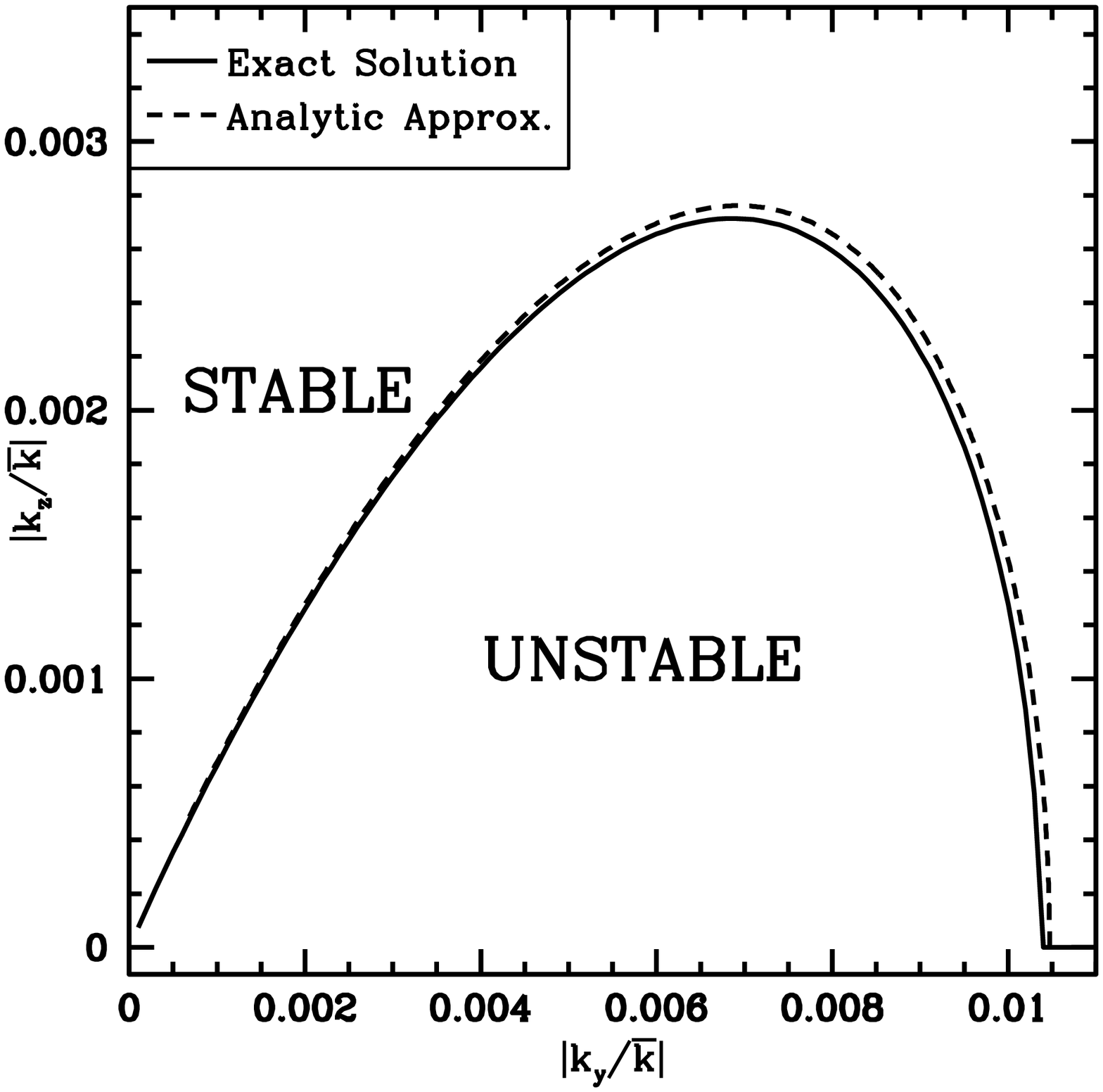}
 \hspace{1cm} \includegraphics[width=.4\textwidth]{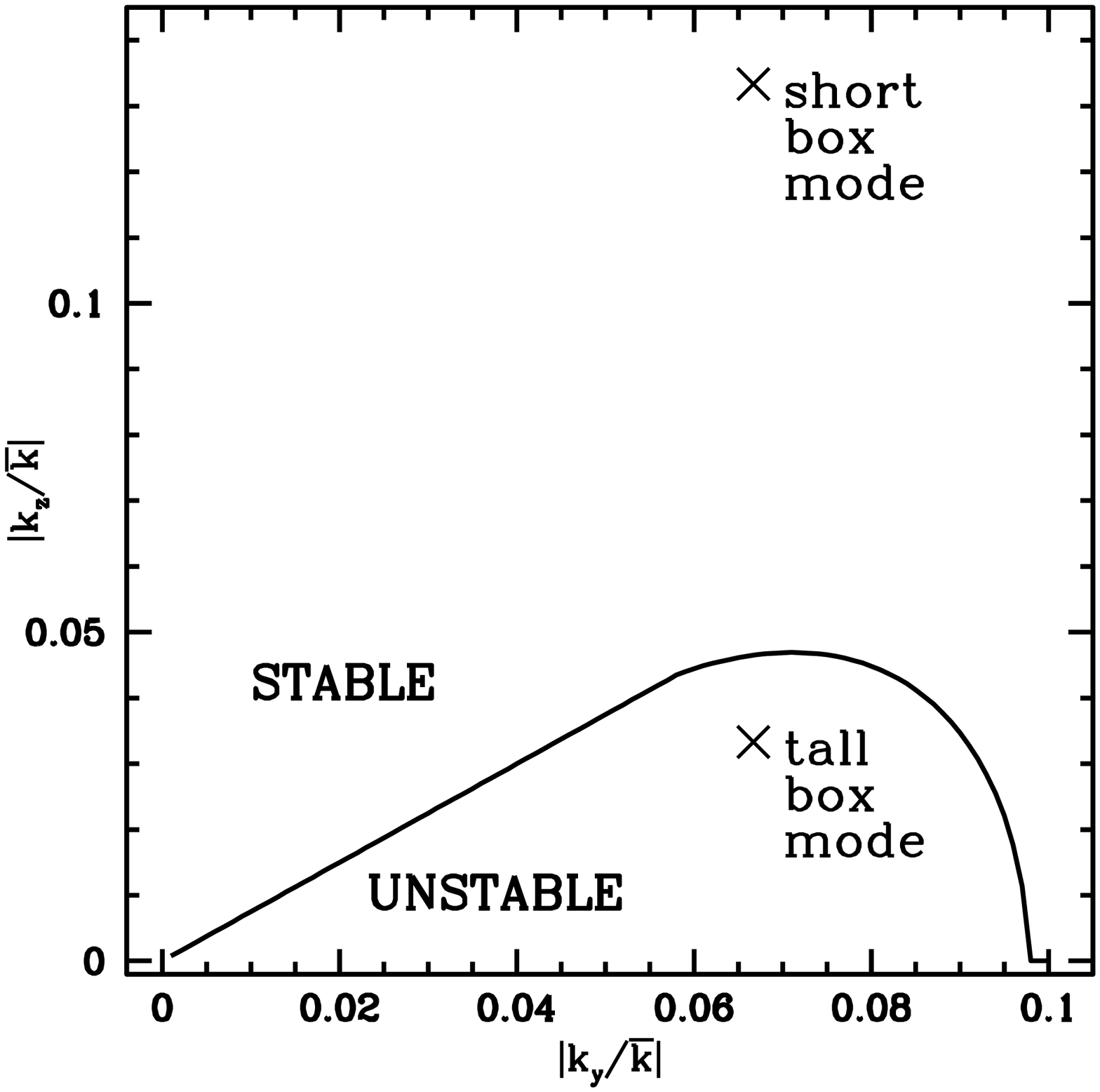}  
  \caption{{\bf Curves of Marginal Stability for a Mother Mode
  With  $\bld{\bar{\omega}=0.005}$ (left panel) and
  $\bld{\bar{\omega}=0.05}$ (right panel):}  Left panel corresponds to Figure \ref{fig:nonlin} 
  and right panel corresponds to the pseudospectral simulations of Figures \ref{fig:shortsim}-\ref{fig:tallsim}.
  We set $\Omega=1$ and $q=3/2$.
To make the solid lines in these plots (the ``exact solutions''), 
we repeated the integrations that produced the lines in Figure \ref{fig:nonlin}, but varying
$k_z$ for each $k_y$ until perturbations neither grew nor decayed.  The dashed line
in the left panel
shows that the analytic approximation of equation (\ref{eq:amp}) agrees reasonably well
with the exact solution.  We do not show equation (\ref{eq:amp}) in the right panel because
the agreement is poorer there. Right panel shows two X's for the values of the
smallest non-zero $|k_y|$ and $|k_z|$ in the simulations of Figures \ref{fig:shortsim}-\ref{fig:tallsim},
i.e., $|k_y/\bar{k}|=L_x/L_y=0.067$ for both simulations, and $|k_z/\bar{k}|=L_x/L_z=0.13$
for the short box and $=0.033$ for the tall box.  The tall box contains a 3D-unstable mode
that leads to the destruction of the vortex into a turbulent-like state.  The short box contains
no such mode, and is stable to 3D perturbations.
 }
  \label{fig:marg}
\end{figure*}

In the Appendix, we solve  equation (\ref{eq:soneq}) analytically to
 derive the amplification factor $\chi$, which is the ratio of the son's amplitude
at any point in its evolution (e.g., when it is radially symmetric), to the father's amplitude at
the same point in its evolution. 
We find 
\be
\chi=-{\bar{\omega}\over q}\bar{\tau}^\delta\sqrt{\pi}{1+\delta\over\delta^2}
\left(
1+{q\Omega\over\kappa^2}(1-\delta)
\right)
{\Gamma(1+\delta/2)\over\Gamma(1/2+\delta/2)}
\label{eq:amp}
\ee
where $\delta=\sqrt{1-4\beta^2}$.
Equation (\ref{eq:amp}) is applicable in the limit $|\bar{\omega}|/q\ll 1$.
For 2D modes ($\beta=0\Rightarrow \delta=1$), it recovers  equation
42  of L07 (see also eq. [\ref{eq:2dinst}] of this paper): 
\be
\chi_{2D}=-\pi{\bar{\omega}\over q}{\bar{k}\over k_y} \ .
\ee

Marginally stable modes have $|\chi|=1$.  
Figure \ref{fig:marg} plots curves of marginal stability. 
The left panel is for the case $\bar{\omega}=0.005$, as in Figure \ref{fig:nonlin}, and the right
panel is for $\bar{\omega}=0.05$, as in the pseudospectral simulations presented 
at the outset of this paper (eq. [\ref{eq:initz}]; Figs. \ref{fig:shortsim}-\ref{fig:tallsim}).
The left panel shows that equation (\ref{eq:amp}) gives a fair reproduction of the exact curve.
We do not show equation (\ref{eq:amp}) in the right panel, because it gives poorer
agreement there (since $|\bar{\omega}|/q$ is too large). In the right panel we also plot X's for the values
of the smallest nonvanishing 
3D wavenumbers in the simulations of Figures \ref{fig:shortsim}-\ref{fig:tallsim}.
In the short-box simulation, all 3D modes lie in the stable zone.  Therefore the dynamics remains
two-dimensional.  But in the 3D box, there is a 3D mode in the unstable zone that destroys the vortex
and gives rise to turbulent-like behavior.

It is interesting to consider briefly how the instability described here connects with the
Rayleigh-unstable case, which occurs when $\kappa^2<0$ . 
  At small $|k_y|$, the  marginally stable curves in Figure \ref{fig:marg}
 are
given by $|\beta|=1/2$, where $\beta=(\kappa/q)(k_z/k_y)$.  Hence if one decreases
$\kappa$ from its Keplerian value $\Omega$, the marginally stable curve becomes
steeper in the $k_z-k_y$ plane, and an increasing number of 3D modes become
unstable.  As $\kappa\rightarrow 0$, if a 2D mode with some $k_y$ is unstable, then
so are all 3D modes with the same $k_y$.  Therefore any 2D-unstable state is also
3D-unstable, and any forming vortex would decay into turbulence.

\section{Conclusions}
\label{sec:conc}

Our main result follows from Figure \ref{fig:marg}, which maps
out the stability of a 
``mother mode'' (i.e., a mode with wavevector $\bar{k}\bld{\hat{x}}$
 and amplitude $\bar{\omega}$)
to nonaxisymmetric 3D perturbations. 
A mother mode is unstable 
provided that the $k_y$ and $k_z$ of the
 nonaxisymmetric perturbations  satisfy both
$|k_y|\lesssim \bar{k}\bar{\omega}/q$ and $|k_z|\lesssim |k_y|$, dropping
order-unity constants.
Based on this result, we may understand the formation, survival, 
and destruction of vortices.
Vortices form out of mother modes that are unstable to 2D ($k_z=0$)
perturbations. Mother modes that are unstable to 2D modes but stable
to 3D ($k_z\ne 0$) ones, form into long-lived vortices.  Mother modes
that are unstable to both 2D and 3D modes are destroyed.
Therefore a mother mode with given $\bar{k}$ and $\bar{\omega}$
will form into a vortex if the disk has a sufficiently large circumferential
extent and a sufficiently small scale height, i.e., if $r\gtrsim \bar{k}^{-1}q/\bar{\omega}$
and $h\lesssim \bar{k}^{-1}q/\bar{\omega}$, where $r$ is the distance
to the center of the disk, and $h$ is the scale height.
Alternatively, 
the mother mode will be destroyed in a turbulent-like state
if  both $r$ and $h$ are sufficiently large
 ($r\gtrsim \bar{k}^{-1}q/\bar{\omega}$
  and
$h\gtrsim \bar{k}^{-1}q/\bar{\omega}$).

Our result
has a number of astrophysical consequences.
In protoplanetary disks that do not contain any vortices,
solid particles drift inward. 
Gas disks orbit at sub-Keplerian speeds,  
$v_{\rm gas}\sim \Omega r (1-\eta)$, where 
$\Omega r$ is the Keplerian speed and $\eta\sim  (c_s/\Omega r)^2$, with
$c_s$  the sound speed.  Since solid particles  would orbit
at the Keplerian speed in the absence of gas, the mismatch of speeds
between solids and gas produces a drag on the solid particles, removing
their angular momentum and causing them to fall into the star.
For 
example, in the minimum mass solar nebula, meter-sized
particles fall in from 1 AU in around a hundred years.  
This rapid infall presents
a serious problem for theories of planet formation, since it is
difficult to produce planets out of dust in under a hundred years.
Vortices can solve this problem \citep{BS95}.
A vortex that has 
excess vorticity $-\bar{\omega}$ 
and radial width $1/\bar{k}$ can halt the infall of particles
provided that $\bar{\omega}/\bar{k}\gtrsim (\Omega r)\eta$, 
because the gas speed induced by such a vortex more than 
compensates for
the sub-Keplerian speed induced by gas pressure.\footnote{
We implicitly assume here that the stopping time of the particle
due to gas drag
is comparable to the orbital time, which is true for meter-sized particles
at 1 AU in the minimum mass solar nebula.  A more careful treatment
shows that
a vortex can stop a particle with stopping time $t_s$ provided
that  $\bar{\omega}/\bar{k}\gtrsim (\Omega r)(\Omega t_s)\eta$ 
\citep{Youdin08}.
}
Previous simulations implied that 3D vortices rapidly 
decay, and so cannot  prevent the rapid infall of solid
particles \citep{BM05b,SSG06}.  Our result
shows that vortices can survive within disks, and so restores
the viability of vortices as a solution to the infall problem.

A more important---and more speculative---application of our result is
to the transport of angular momentum within neutral accretion disks. 
In our simulation of a vortex in a tall box, we found that as the vortex 
decayed it transported angular momentum outward at a nearly
constant rate for hundreds of orbital times.
 If decaying vortices transport a significant amount of angular momentum
in disks, they would resolve one of the most important outstanding questions
in astrophysics today: what causes hydrodynamical accretion disks to accrete?
To make this speculation more concrete, one must understand the amplitude
and duration of the ``turbulence'' that results from decaying vortices.  
This is a topic for future research.

In this paper, we considered only the effects of rotation and shear on
the stability of vortices, while we neglected the
effect of vertical gravity.  
There has been a lot of research in the geophysical community 
on the dynamics of fluids in the presence of vertical gravity, since 
stably stratified fluids are very common on Earth---in the atmosphere, oceans, 
and lakes.  In numerical and laboratory experiments of strongly stratified
flows, thin horizontal ``pancake vortices'' often form, and fully developed 
turbulence is characterized by thin horizontal layers. 
\citep[e.g., ][]{BBLC07}.  Pancake vortices are stabilized by vertical gravity,
in contrast to the vortices studied in this paper which are stabilized by rotation.
Gravity inhibits vertical motions because of buoyancy: it costs gravitational energy
for fluid to move vertically.
The resulting quasi-two-dimensional
flow can form into a vortex.\footnote{\cite{BC00} 
show that
a vertically uniform
vortex column in a stratified (and non-rotating and non-shearing) 
 fluid suffers an instability (the ``zigzag instability'') that is 
characterized by a typical vertical lengthscale $\lambda_z\sim U/N$, where $U$ is 
the horizontal speed induced by the vortex, $N$ is the Brunt-V\"ais\"al\"a frequency,
and the horizontal lengthscale of the vortex $L_h$ is assumed to be much
greater that  $\lambda_z$ (hence the pancake structure).
 We may understand
Billant \& Chomaz's result in a crude fashion with an argument similar to that employed in the introduction
to explain the destruction of rotation-stabilized vortices: since the frequency
of buoyancy waves is $Nk_x/k_z$ (when $|k_x|\ll |k_z|$), and since the frequency
at which fluid circulates around a vortex is $U/L_h\sim k_xU$, there is a resonance between
these two frequencies for vertical lengthscale $1/k_z\sim U/N$.} 
We may speculate that in an astrophysical disk vertical gravity provides an 
additional means to stabilize vortices, in addition to rotation.  
But to make this speculation concrete, the theory presented in this paper
should be extended to include vertical gravity.

We have not addressed in this paper the origin of the axisymmetric structure
(the mother modes) that give rise to surviving or decaying vortices.
One possibility is that decaying vortices can produce more axisymmetric
structure, and therefore they can lead to self-sustaining turbulence.
This seems to us unlikely.
We have not seen evidence for it in our simulations, but this could
be  because
of the modest resolution of our simulations.  
Other possibilities for the generation of axisymmetric structure include thermal instabilities, such
as the baroclinic instability, or convection, or
stirring by planets.  This, too, is a topic for future
research.

\appendix
\section{Analytic Expression for Growth Factor $\chi$}

In this Appendix, we derive equation (\ref{eq:amp}) by analytically
integrating equation (\ref{eq:soneq}) for the son's vorticity, given the father's vorticity
as a function of time (\S \ref{sec:lin}),
and taking the mother's vorticity $\bar{\omega}$
 to be constant, which is valid when the father's amplitude is 
small relative to the mother's.
The numerical integral of equation (\ref{eq:soneq})  is shown in Figure \ref{fig:nonlin}.
Recall that initially  $\tau=\bar{\tau}>0$ and $\tau$ decreases in time, and typically
$\bar{\tau}\gg 1$.
It simplifies the analysis to work with the son's ``normal-mode'' amplitudes $\omega_A'$ and
$\omega_B'$, defined from $\omega_x'$ and $\omega_{yz}'$ via 
(eqs. [\ref{eq:oxeq}] and [\ref{eq:powerlaw}])
\be
\label{eq:sonnm}
\left(\begin{array}{c}{\omega}_x' \\ {\omega}_{yz}'\end{array}\right)
=
\left(\begin{array}{cc}
{\kappa\over 2\beta\Omega}{1-\delta\over 2\tau'}|\tau'|^{1-\delta\over 2} & 
{\kappa\over 2\beta\Omega}{1+\delta\over 2\tau'}|\tau'|^{1+\delta\over 2} 
 \\
|\tau'|^{1-\delta\over 2}   
  & 
|\tau'|^{1+\delta\over 2}  
  \end{array}\right)
\left(\begin{array}{c}{\omega}_A' \\ {\omega}_B'\end{array}\right) \ .
\ee
where
\be
\tau'\equiv\tau+\bar{\tau}
\ee
Substituting this into equation (\ref{eq:soneq}), the time derivative of the above
matrix cancels the homogeneous term in that equation if we approximate
$1+\tau^{'2}\simeq  \tau^{'2}$, which holds until just before the time
that $\tau=-\bar{\tau}$.  The inhomogeneous term produces
\beqn
{d\over d\tau}\omega_A'={\bar{\omega}\over q}{1\over\delta}|\tau'|^{1+\delta\over 2}
\left(
\omega_{yz}
\left[
{1\over 1+\tau^2}
\left(
{2\beta^2q\Omega\over\kappa^2}+{1+\delta\over 2}{\bar{\tau}\over\tau'}
\right)
-{1+\delta\over 2}{1\over \bar{\tau}\tau'}
\right]
+\omega_x{2\beta\Omega\over\kappa}{1\over{\bar{\tau}}}
\right) \ \ , \ \ \tau\gtrsim -\bar{\tau}
\label{eq:aprime}
\eeqn
Since $\omega_x$ and $\omega_{yz}$ are known (\S \ref{sec:lin}), 
a straightforward integration yields $\omega_A'$ just before $\tau=-\bar{\tau}$.
To perform this integral, we resort to some approximations, guided by the solution
shown in Figure \ref{fig:nonlin}.
For the first term, we need
\beqn
\int_{\bar{\tau}}^{-\bar{\tau}}\omega_{yz}{(\tau+\bar\tau)^{1+\delta\over 2}\over 1+\tau^2}d\tau
&\approx&
-{1\over \beta^2}\int_{\bar{\tau}}^{-\epsilon\bar\tau}
{d^2\omega_{yz}\over d\tau^2}(\tau+\bar\tau)^{1+\delta\over 2}
d\tau
+
\omega_B\int_{-\epsilon\bar\tau}^{-\bar\tau}
|\tau|^{{1+\delta\over 2}-2}
(\tau+\bar\tau)^{1+\delta\over 2}
d\tau 
\\
&\approx&
\omega_B\bar{\tau}^\delta{1+\delta\over 2\beta^2}\epsilon^{-1+\delta\over 2}
-\omega_B\bar{\tau}^\delta\int_{\epsilon}^1s^{-3+\delta\over 2}(1-s)^{1+\delta\over 2}ds\\
&\approx&
\omega_B{\bar\tau}^\delta2^{-\delta}\sqrt{\pi}{1+\delta\over1-\delta}
{\Gamma(1/2+\delta/2)\over \Gamma(1+\delta/2)}
\eeqn
where $\epsilon$ is a parameter that satisfies $1\gg \epsilon\gg 1/\bar\tau$.  In the first line,
we used equation (\ref{eq:bh}),
and
we discarded the $\omega_A$ mode from the second integral 
because the $\omega_B$ mode increases faster
with increasing $|\tau|$. 
From Figure \ref{fig:nonlin}, $\omega_A'$ nearly vanishes until $\tau\simeq 0$.  Therefore,
in the second line we approximated the first integral as 
$-(\bar{\tau}^{1+\delta\over 2}/\beta^2){d\omega_{yz}/d\tau}\vert_{-\epsilon}$.
The third line holds in the limit of small $\epsilon$.
The other three terms in equation (\ref{eq:aprime}) are integrated similarly, yielding
\be
\omega_A'=\omega_B{\bar{\omega}\over q}{1\over\delta}{\bar\tau}^\delta 2^{-\delta-1}\sqrt{\pi}
{\Gamma(1/2+\delta/2)\over \Gamma(1+\delta/2)}
\left(
1+{q\Omega\over\kappa^2}(1-\delta)
\right){(1+\delta)^2\over 1-\delta}
\label{eq:omap}
\ee
just before the time when $\tau=-\bar{\tau}$, i.e., just before the son is radially symmetric. 
At this time, Figure \ref{fig:nonlin} shows that
$\omega_B'$ very nearly vanishes.  Therefore, just after the son is radially symmetric, 
it will have $\omega_B'=T_{BA}{\omega_A'}$ (eq. [\ref{eq:trans}]), with $\omega_A'$
given by equation (\ref{eq:omap}). This gives the amplification factor $\chi\equiv\omega_B'/\omega_B$
that is displayed in equation (\ref{eq:amp}).

\bibliographystyle{apj}

\end{document}